\documentclass[journal]{IEEEtran}
\usepackage{amsmath,amsfonts}
\usepackage{multirow}%
\usepackage{algorithm}%
\usepackage{algorithmicx}%
\usepackage{algpseudocode}%
\usepackage{array}
\usepackage[caption=false,font=footnotesize,labelfont=rm,textfont=rm]{subfig}
\usepackage{textcomp}
\usepackage{stfloats}
\usepackage{url}
\usepackage{verbatim}
\usepackage{graphicx}
\usepackage{listings}%
\usepackage{cite}
\usepackage{booktabs}
\usepackage{subfig}
\usepackage{threeparttable}
\usepackage{makecell}
\usepackage{doi}
\usepackage{pifont}
\usepackage{xcolor}

\begin{document}
 
\title{\huge{Factorized Disentangled Representation Learning for Interpretable Radio Frequency Fingerprint}}
\author{Yezhuo Zhang, Zinan Zhou, Guangyu Li and Xuanpeng Li
\thanks{This work has been submitted to the IEEE for possible publication.
 Copyright may be transferred without notice, after which this version may
 no longer be accessible. (Corresponding author: Xuanpeng Li.)}
\thanks{Yezhuo Zhang, Zinan Zhou and Xuanpeng Li are with the School of Instrument Science and Engineering, Southeast University, Nanjing, 210096, Jiangsu, China (e-mail: zhang\_yezhuo@seu.edu.cn; zhouzinan919@seu.edu.cn; li\_xuanpeng@seu.edu.cn).}
\thanks{Guangyu Li is with the School of Computer Science and Engineering, University of Science and Technology, Nanjing, 210094, Jiangsu, China (email: guangyu.li2017@njust.edu.cn).}
\thanks{Digital Object Identifier XX.XXXX/LCOMM.2024.XXXXXXX}
}

\markboth{Journal of \LaTeX\ Class Files,~Vol.~14, No.~8, August~2021}%
{Shell \MakeLowercase{\underline{et al.}}: A Sample Article Using IEEEtran.cls for IEEE Journals}

\IEEEpubid{0000--0000/00\$00.00~\copyright~2021 IEEE}

\maketitle
\begin{abstract}

  In response to the rapid growth of Internet of Things (IoT) devices and rising security risks, Radio Frequency Fingerprint (RFF) has become key for device identification and authentication. However, various changing factors—beyond the RFF itself—can be entangled from signal transmission to reception, reducing the effectiveness of RFF Identification (RFFI). Existing RFFI methods mainly rely on domain adaptation techniques, which often lack explicit factor representations, resulting in less robustness and limited controllability for downstream tasks. To tackle this problem, we propose a novel Disentangled Representation Learning (DRL) framework that learns explicit and independent representations of multiple factors, including the RFF. Our framework introduces modules for disentanglement, guided by the principles of explicitness, modularity, and compactness. We design two dedicated modules for factor classification and signal reconstruction, each with tailored loss functions that encourage effective disentanglement and enhance support for downstream tasks. Thus, the framework can extract a set of interpretable vectors that explicitly represent corresponding factors. We evaluate our approach on two public benchmark datasets and a self-collected dataset. Our method achieves impressive performance on multiple DRL metrics. We also analyze the effectiveness of our method on downstream  RFFI task and conditional signal generation task. All modules of the framework contribute to improved classification accuracy, and enable precise control over conditional generated signals. These results highlight the potential of our DRL framework for interpretable and explicit RFFs.

\end{abstract}

\begin{IEEEkeywords}
Radio frequency fingerprint, disentangle representation learning, specific emitter identification, domain adaptation.
\end{IEEEkeywords}

\section{Introduction}\label{sec:intro}
\subsection{Background}\label{background}
\IEEEPARstart{I}{n} recent years, the rapid growth of the Internet of Things (IoT) has been fueled by advancements in wireless communication, big data, and artificial intelligence, which have led to the deployment of an increasing number of IoT devices across various scenarios \cite{Nguyen_6G_2022}. While IoT offers significant benefits, it poses substantial security concerns. The open nature of IoT makes it vulnerable to malicious attacks, especially because the communication occurs via wireless signals. Unauthorized spectrum usage by malicious devices further compromises IoT security and disrupts the efficient utilization of spectrum resources \cite{xu_device_2016,alwarafy_survey_2021}. There is an urgent need for reliable and efficient device identification methods that offer both high security and low complexity to mitigate these risks \cite{xiao_phy_2018,Hamdaoui_deep_2021}.

Radio Frequency Fingerprint Identification (RFFI) has emerged as a key technology for device identification and authentication. By exploiting hardware-induced imperfections in transmitted signals, RFFI enables the extraction of unique physical-layer signatures, termed Radio Frequency Fingerprints (RFF), which are inherently tied to the manufacturing variations of Radio Frequency (RF) components such as oscillators, power amplifiers, and mixers \cite{liu_Machine_2021,huan_Carrier_2023}. This technology holds significant promise in military applications for threat detection and in civilian domains for enhancing IoT security against spoofing attacks \cite{jagannath_comprehensive_2022,meng_Survey_2024}. 

\begin{figure}[t]
  \centering
  {\includegraphics[width=1\columnwidth]{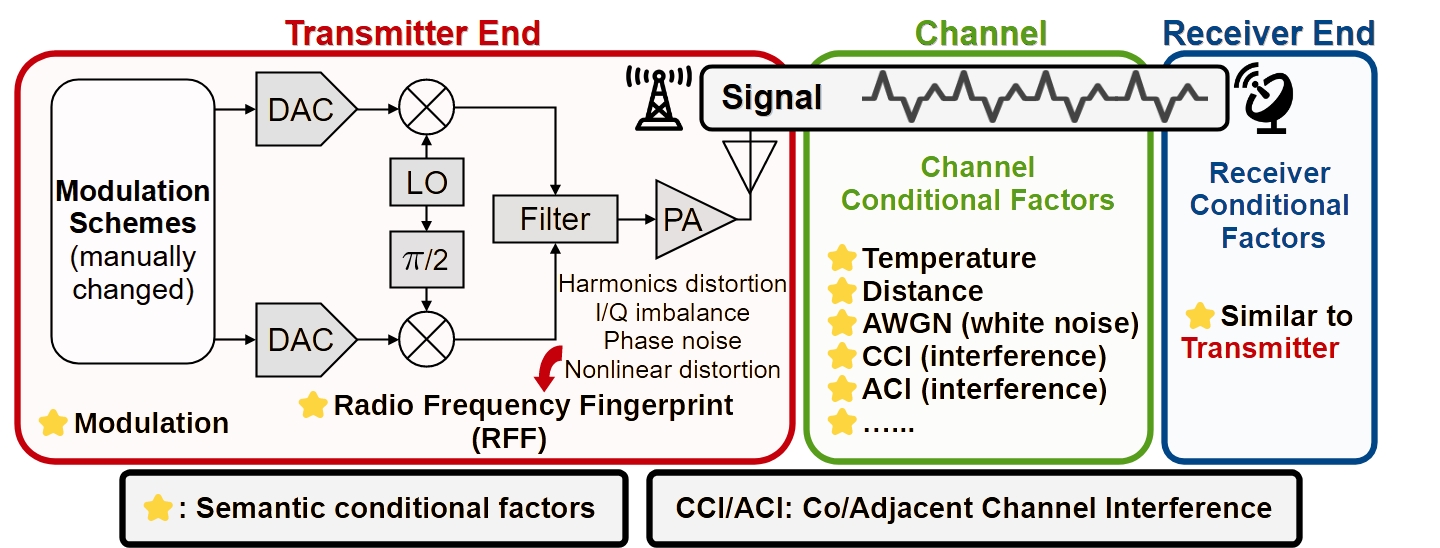}\hspace{5mm}
  \caption{Semantic conditional factors that occur during the process of signal emission, transmission, and reception.}\label{fig:factors}}
\end{figure}

However, the practical deployment of RFFI faces critical challenges due to distribution shifts caused by multiple semantic conditional factors. As depicted in Fig. \ref{fig:factors}, factors include the RFF, modulation variations, dynamic channel conditions, and receiver hardware discrepancies, etc. \cite{meng_Survey_2024,xie_Radio_2024}. These factors degrade the generalization of Deep Learning (DL)-based RFFI methods. It is necessary to have disentangled representations that are interpretable and explicit to improve the performance of the task.

\IEEEpubidadjcol{}


\subsection{Related Works}\label{ralated_works}

The processes of RFFI can be broadly categorized into Manually Extracted Feature-based, DL-based, DL-based with Domain Adaptation (DA), and the proposed Disentangled Representation Learning (DRL)-based approaches.

\subsubsection{Manual Feature Extraction Approaches}

Manual features include Carrier Frequency Offset (CFO), In-phase/Quadrature (I/Q) imbalance, modulation parameters, power amplifier nonlinearity, and signal envelope characteristics, etc. \cite{he_radio_2023,  sun_radio_2022, li_radio_2022}. Common classification models for these features include Support Vector Machines (SVM) and K-Nearest Neighbors (KNN). For example, He et al. \cite{he_radio_2023} proposed an RFFI framework addressing distortions caused by multipath fading, CFO, and phase offset, using SVM to classify features and leveraging frequency correlation for enhanced accuracy and robustness. However, manually extracted features vary across signal sources, making them sensitive to noise and environmental variations,  potentially fail to capture all critical information within the RFF, limiting RFFI accuracy due to the small feature space.

\subsubsection{Deep Learning-Based Approaches}

DL-based methods automatically extract features and directly process signals to output transmitter labels. These end-to-end models have been widely adopted, with approaches such as Shen et al. \cite{shen_radio_2021} using spectrogram and Convolutional Neural Networks (CNN) for Long Range (LoRa) systems, and He et al. \cite{he_cooperative_2020} combining Support Vector Machine (SVM) and Long Short-Term Memory (LSTM). The introduction of ResNet \cite{pan_specific_2019, wang_radio_2020} has led to models like DRSN \cite{zhao_deep_2020} and CVCNN \cite{wang_efficient_2021} that further address noise and I/Q signals. However, DL-based methods face challenges such as the high cost and time required to acquire labeled data and the lack of interpretability due to their "black-box" nature.

\subsubsection{Domain Adaptation in RFFI}

It has become recognized that ideal experimental conditions lead to better outcomes, but achieving such conditions in real electromagnetic environments is challenging. This issue in RFFI is attributed to \textit{distribution shifts}, where a model trained on one distribution performs poorly on another. Variations from receivers, modulation schemes, channel conditions, and temperature changes are considered distribution shifts, which have led to the exploration of Domain Adaptation (DA) in RFFI \cite{zha_Crossreceiver_2023,wan_Variablechannel_2023,zhang_Variablemodulation_2022,saeif_Dayaftertomorrow_2023}.

DA methods measures \textit{domain discrepancies} and aligns domains to reduce differences in the representation space. Receiver domain shifts have received the most attention \cite{bao_Receiveragnostic_2023,zha_Crossreceiver_2023,yang_Mitigating_2024,shen_Receiveragnostic_2023}. Adversarial training methods, such as Domain-Adversarial Neural Networks (DANN), learn receiver-agnostic features by minimizing global domain discrepancies \cite{zhang_Variablemodulation_2022}. Subdomain adaptation techniques, like Local Maximum Mean Discrepancy (LMMD), aligns conditional distributions of shared classes \cite{bao_Receiveragnostic_2023,zha_Crossreceiver_2023}. Hybrid approaches combining unsupervised DA with semi-supervised fine-tuning have also gained popularity \cite{wan_Variablechannel_2023,wan_VCSEI_2024,zhang_Semisupervised_2024}, leveraging a few labeled target samples to refine domain-invariant features and enhance model adaptability. This paradigm has been applied successfully in channel-variant scenarios. Wan et al. extracted device-specific components using Global Semantic Consistency (GSC) to mitigate channel variation impacts \cite{wan_VCSEI_2024}, while Pan et al. achieved similar results using correlation alignment (CORAL) \cite{pan_Equalization_2024}.

While DA methods improve RFFI robustness by addressing distribution shifts, they mainly focus on one domain other than RFF, and lack explicit domain representations.

\subsubsection{Attention to Factors in RFFI}

As the concept of ``domain'' has gradually gained attention in the RFFI field, a few researchers have recognized that disentanglement between  ``factors'' is key to addressing distribution shifts. Elmaghbub et al. \cite{elmaghbub_ADLID_2023} proposed an unsupervised DA framework, ADL-ID, employing an adversarial network to separate the feature vector into RFF and domain-specific factors, using only the RFF component for classification. Yin et al. \cite{yin_FDSNet_2022} introduced the Feature Domain Separation Network (FDSNet), which embeds raw signals into separate spaces for RFF's information and other factors' information. To address the issue of receiver information entangled in RFF, Liu et al. \cite{liu_Receiveragnostic_2023} proposed a receiver-independent model. Xie et al. \cite{xie_Disentangled_2023} separated signals into device-relevant and device-irrelevant classification via adversarial learning. Although researchers have recognized the need to distinguish between different factors, the number of factors examined remains limited to two, and there is still no unified framework for disentangling the entangled factors.

\subsection{Challenge and Motivation}
To sum up, significant challenges remain unresolved in the aforementioned RFFI methods handling factors entanglement:
\begin{enumerate}
  \item Interpretability: These methods cannot explain or intervene  factors. In other words, different factors involved in signal generation are not explicitly represented as vectors containing semantic information.
  \item Factor Richness: The variable domain in real-world electromagnetic environments are much more complex. Most existing approaches in RFFI focus on adaptation between two specific factors (one RFF and one additional factor). 
  \item Task Limitation: Previous studies have primarily validated model interpretability through classification tasks. In fact, interpretable modules can be applied to a wider range of downstream tasks, such as conditional generation, which provides a more intuitive demonstration of interpretability.
\end{enumerate}

In fact, DRL, as a well-established paradigm, has been widely applied in various fields requiring the handling of complex factors. It ensures that each factor possesses interpretable semantic information corresponding to specific dimensions in the representation space, thereby enhancing the model's performance on downstream tasks \cite{carbonneau_Measuring_2022,rauker_Transparent_2023}.

In Natural Language Processing (NLP), DRL can disentangle grammatical and semantic information in text, facilitating machine translations \cite{Rezende_variational_2015}. In medical image analysis, DRL separates pathological regions, healthy tissues, and other relevant features, providing more interpretable representations for diagnosis \cite{Liu_learning_2022}. The most prominent application of DRL is image generation, where it disentangles factors like pose, background, and object categories \cite{bengio_learning_2009}. This enables the model to control factors independently, resulting in more diverse image generation.

For applying DRL in RFF extraction, the core objective is to disentangle several latent vectors as ``representations'' to capture interpretable and independent factors of variation in the data (e.g., RFF, modulation, channel condition, or receiver's fingerprint). 

\subsection{Contributions}

This paper proposes a DRL-based framework to extract and disentangle the factors—including RFF—that influence signal generation, and applies the representations of these factors to classification and conditional generation tasks. Specifically, modules are designed with different functions and corresponding loss terms from three DRL principles: \textit{explicitness}, \textit{modularity}, and \textit{compactness} (further discussed in Section \ref{subsec:DRL}). Comprehensive experiments are conducted on two public RFFI datasets and one self-collected dataset, demonstrating that the proposed DRL framework effectively disentangles sementic representations. Each module contributes positively to the improvement of classification accuracy, and the representations are capable of guiding the model to generate diverse, factor-controlled signals. The main contributions of this paper are summarized as follows:
\begin{itemize}
  \item As far as we are aware, this study is the first attempt in the field of RFF to explore the impact of multiple factors across the three stages of signal information—transmitters, channels, and receivers. This study leverages DRL to extract separated, explicit representations that exhibit discriminative capability for downstream applications. 
  
  \item We propose a novel neural network framework for communication signal representation. Based on the principles of modularity, explicitness and compactness, we design multiple loss functions to guide the model to obtain explicit representations. Extensive comparative experiments demonstrate that the obtained representations have clear semantic information, each reflects the corresponding factor.
  
  \item The extracted representations improve the performance of downstream tasks, including classification and conditional generation. Experimental results indicate that disentanglement improves classification accuracy and enables customization of conditioned factorized signals.
\end{itemize}

The remainder of the paper is organized as follows: Section \ref{sec:preliminary} describes the DRL problem of communication signals and explain the role of Diffusion Probabilistic Model (DPM) in the DRL process. Section \ref{sec:model} presents the composition of the proposed model, detailing the functions of each component and the design of the loss function. Section \ref{sec:exp_setup} details the datasets and implementation parameters. Section \ref{sec:exp} provides an extensive experimental evaluation, assessing both the quality of the representations and the performance on downstream classification and generation tasks. Section \ref{sec:discussion} offers a discussion on the challenges encountered in disentangling and the issues identified throughout the study. Finally, Section \ref{sec:conclusion} concludes the paper.

\section{Preliminary}\label{sec:preliminary}
\subsection{Problem Statement}

According to the sequence of signal emission, transmission, and reception, the signal data is ultimately captured and stored in a computer following the process shown in Fig. \ref{fig:factors} through the \textbf{Transmitter End}, \textbf{Channel}, and \textbf{Receiver End}. The signal's characteristics are influenced by numerous intentional and unintentional semantic and conditional factors. 

Initially, the signal is modulated according to a predefined modulation scheme, and then passes through hardware modules such as the Digital-to-Analog Converter (DAC), I/Q modulator, oscillator, Power Amplifier (PA), and antenna within the transmitter. Impairments from different hardware components, such as I/Q imbalance and phase noise, lead to a unique ``fingerprint'' to the transmitter. After the signal is transmitted through the antenna into the channel, the channel's characteristics become highly complex. In addition to the transmission distance and temperature, which have previously been examined by researchers \cite{saeif_Dayaftertomorrow_2023}, the channel's transmission characteristics at different frequency bands are also crucial. This includes phenomena such as co-channel interference (CCI) and adjacent-channel interference (ACI) for specific signals. The receiver, conceptually functioning as a ``reverse'' of the transmitter, comprises similar components to those of the transmitter and therefore exhibits a unique receiver-specific RFF.

\begin{figure*}[t]
  \centering
  \subfloat[ ]{\includegraphics[height=.4\columnwidth]{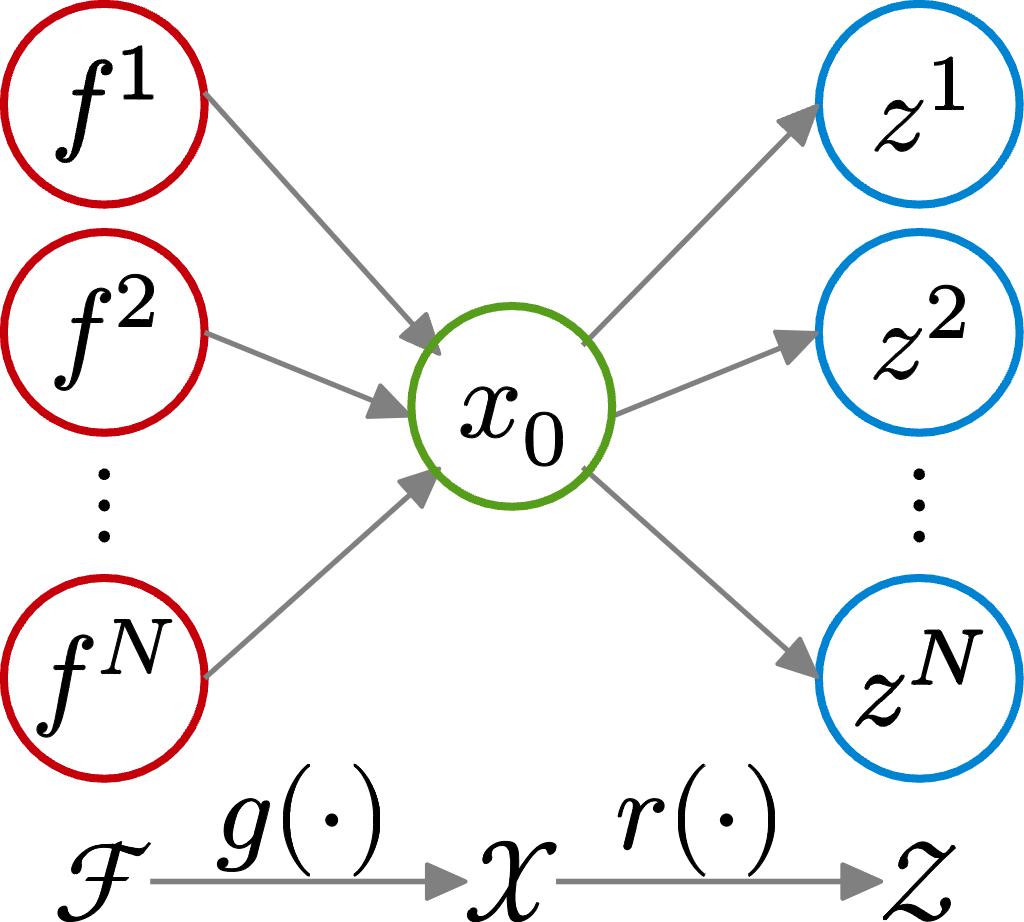}\hspace{8mm}\label{PGM}}
  \subfloat[]{\includegraphics[height=.5\columnwidth]{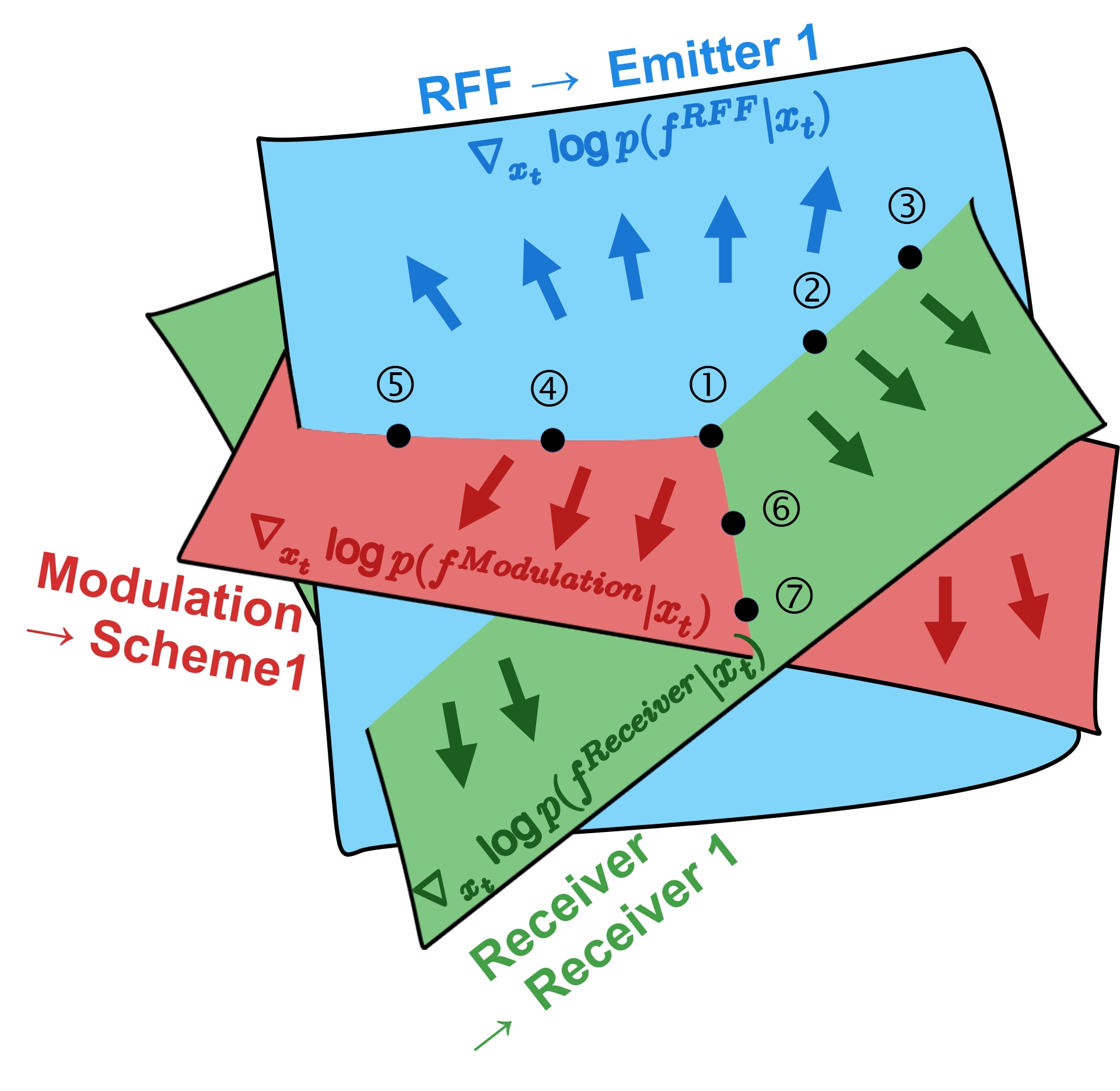}\hspace{8mm}\label{cross_space}}
  \subfloat[]{\includegraphics[height=.5\columnwidth]{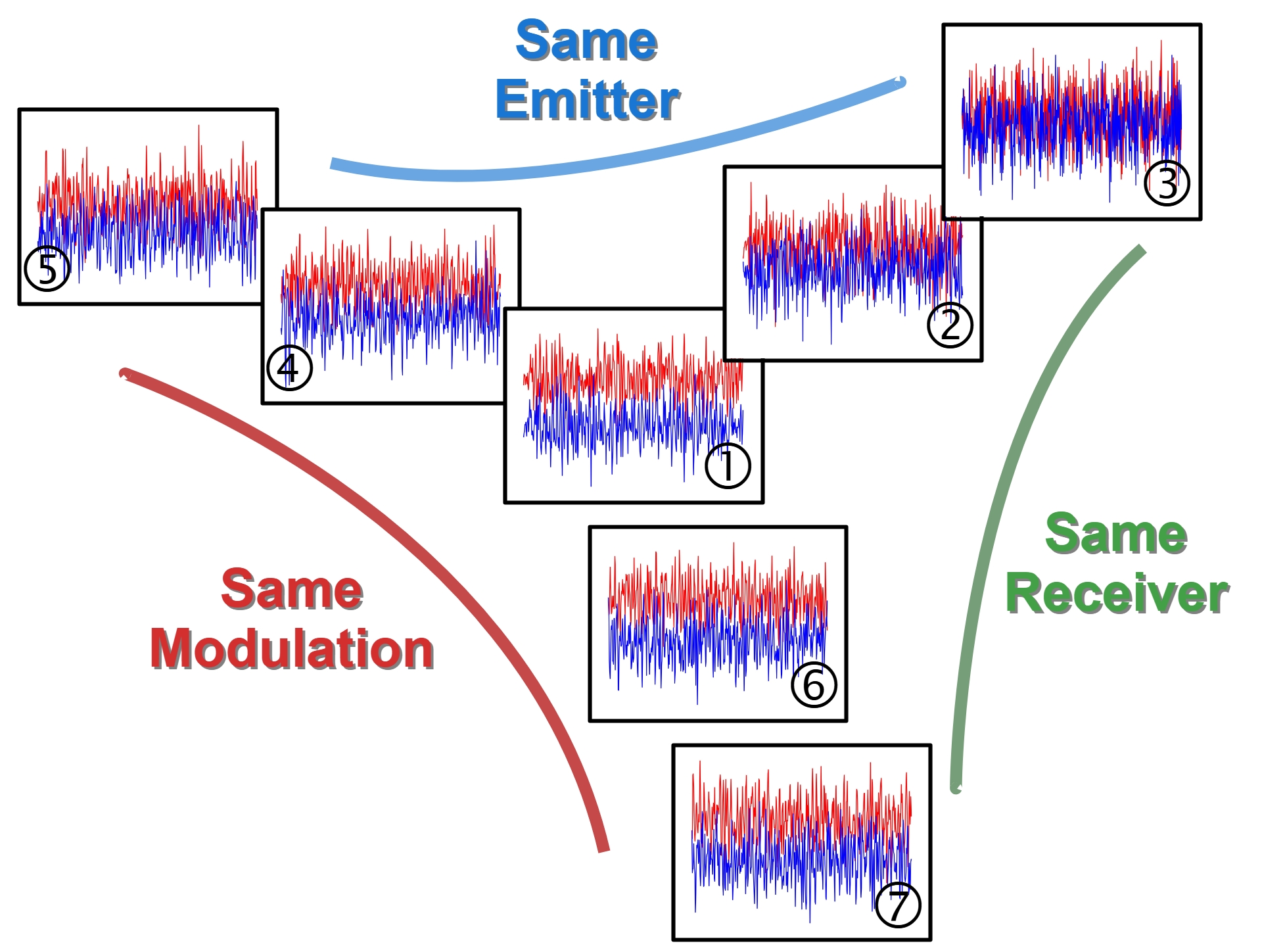}\label{inference_time_length}}
  \caption{Diagram and phenomenon explanation of DRL. 
  (a) shows the Probabilistic Graphical Model (PGM) composed of factor, signal, and representation. (b) demonstrates the feature space distribution \ding{172} $\sim$ \ding{178} in (c) using RFF, Modulation, and Receiver ID as examples. The overlapping surfaces formed by different factors result in signal samples with distinct representations.}\label{cross_space_2}
\end{figure*}

\begin{figure}[t]
  \centering
  \subfloat[]{\includegraphics[width=.3\columnwidth]{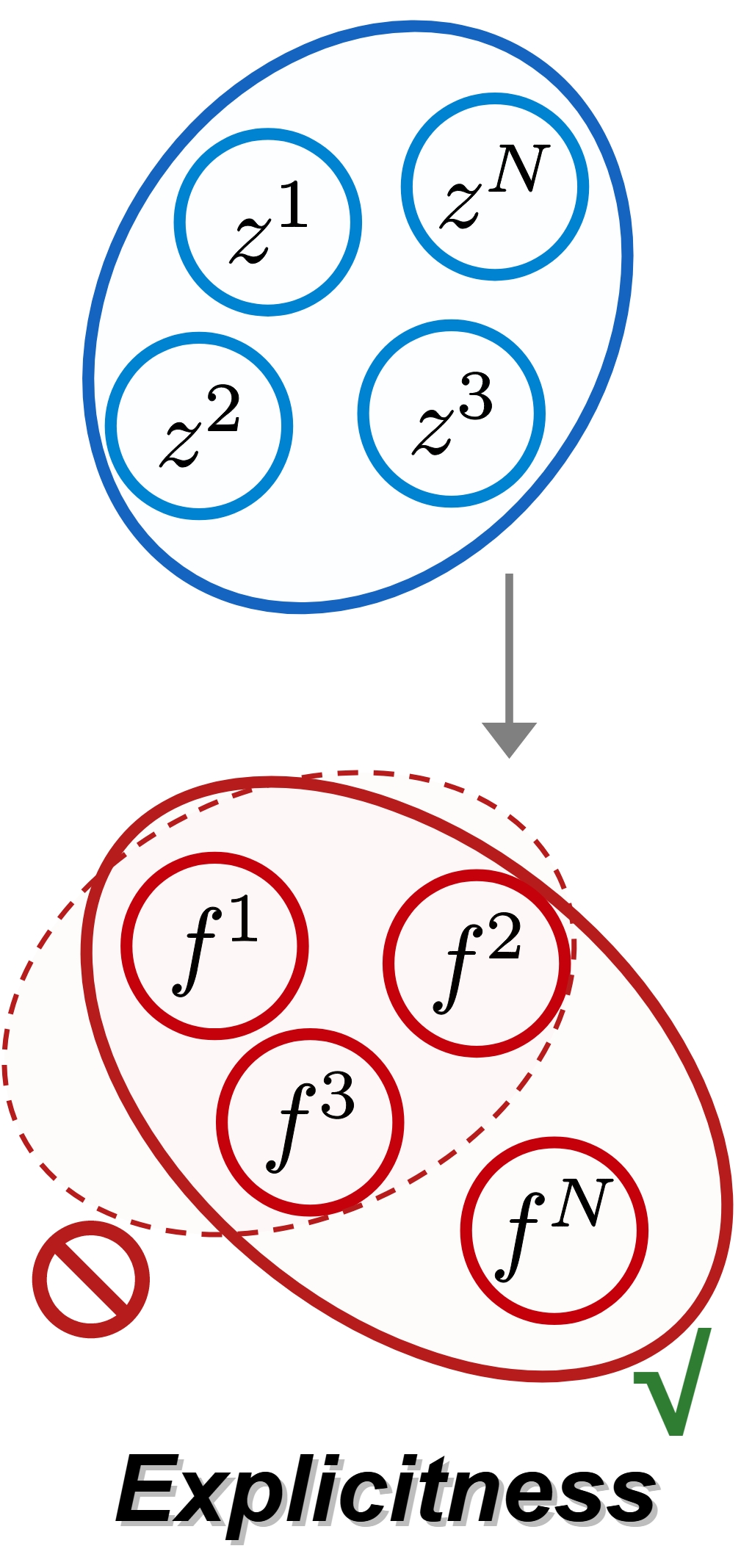}\hspace{1mm}\label{fig:exp}}
  \subfloat[]{\includegraphics[width=.3\columnwidth]{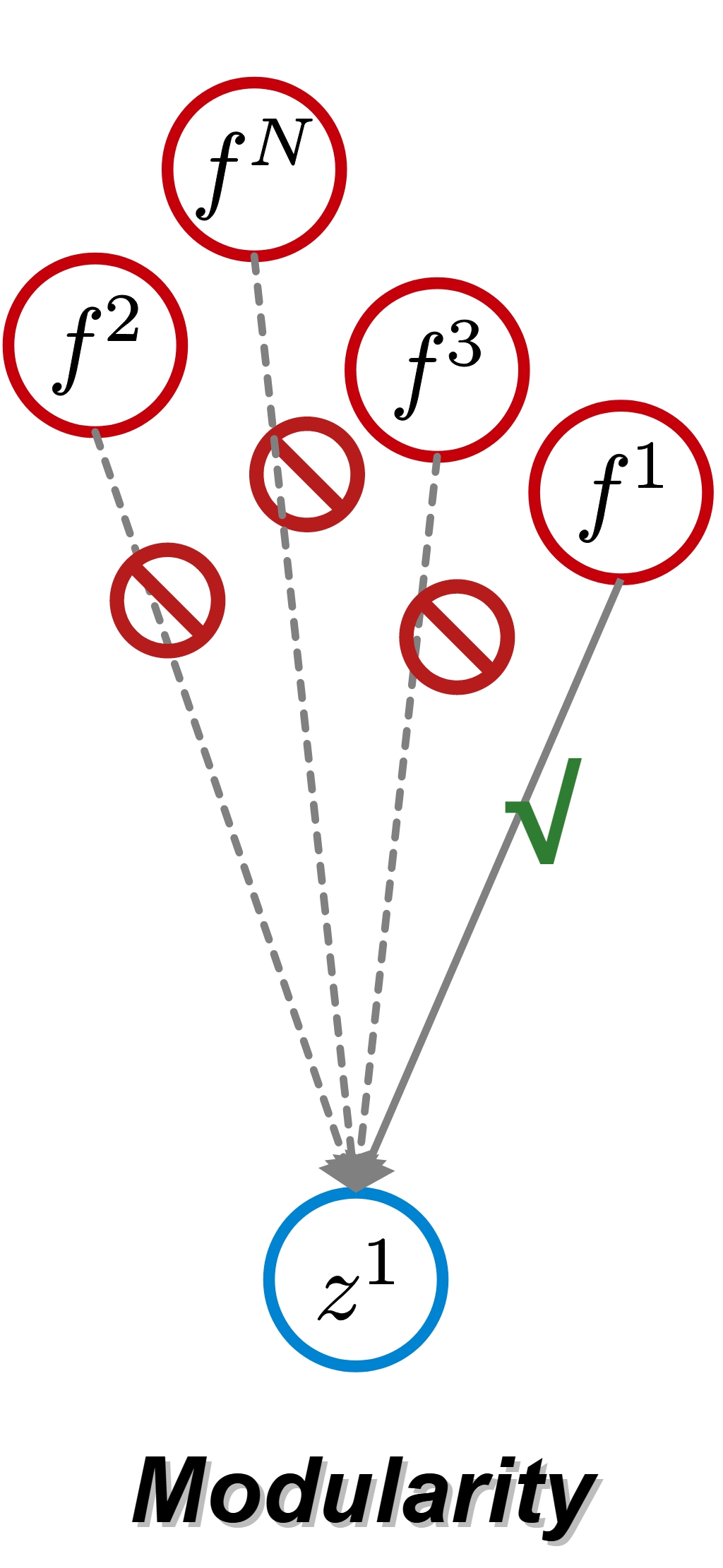}\hspace{1mm}\label{fig:mod}}
  \subfloat[]{\includegraphics[width=.3\columnwidth]{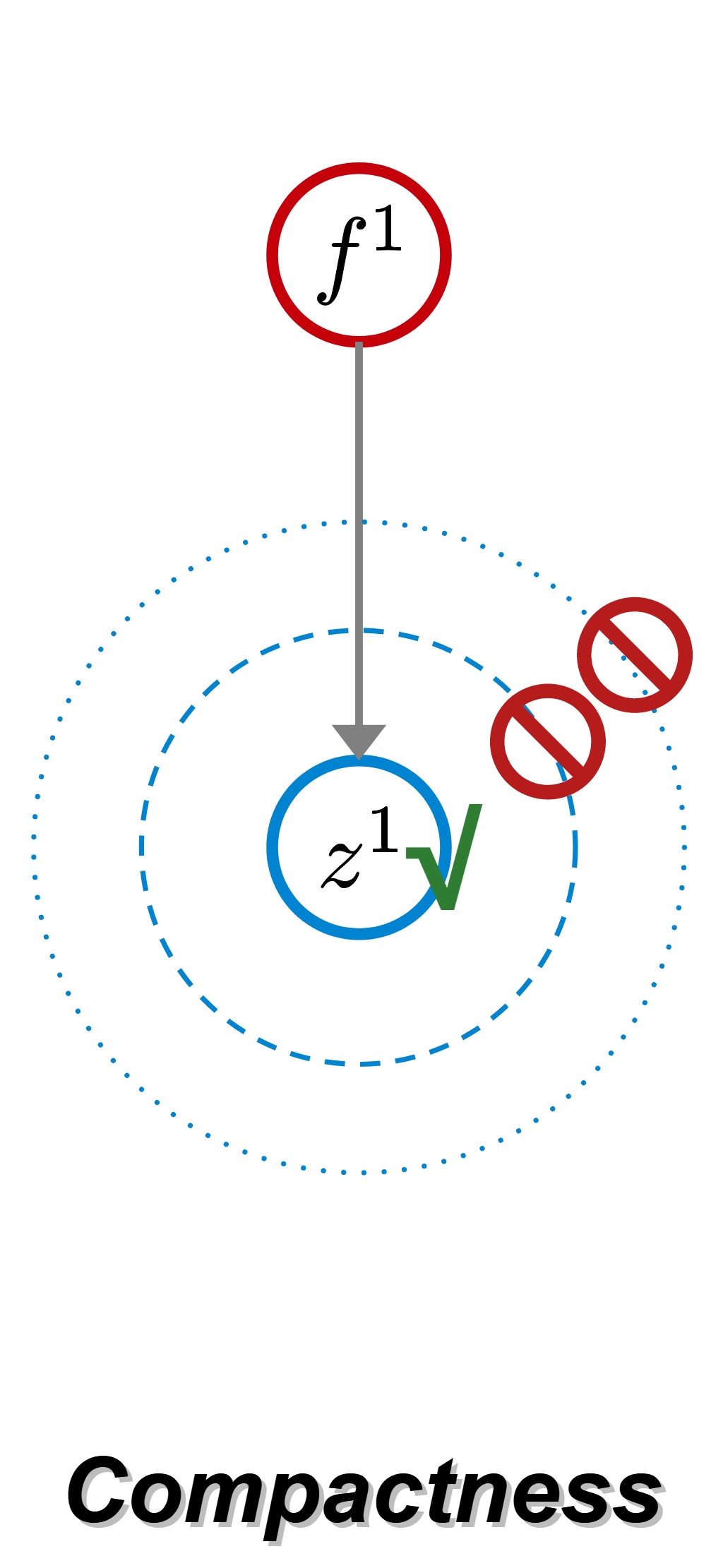}\label{fig:com}}
  \caption{Principles of explicitness, modularity and compactness in DRL. }\label{fig:exp_mod_com}
\end{figure}

Assume that the original signal data $\mathcal{X}=\{x_0 | x_0 \sim p(x_0)\}$ is generated by $N$ underlying ground truth factors $\mathcal{F} = \{f^c | c \in \mathcal{C} = \{1, 2, \ldots, N\}\}$, i.e., $g: \mathcal{F} \mapsto \mathcal{X}$, with the data distribution $p(x_0)$. This implies that the underlying factors condition each sample. Moreover, each factor $f^c$ follows the distribution $p(f^c)$, with $p(f^c)$ representing the distribution of the factor $f^c$. As depicted in Fig. \ref{PGM}, a representation learning algorithm is a mapping $r: \mathcal{X} \mapsto \mathcal{Z}$, where $z \in \mathbb{R}^d$ represents a point in the learned representation space $\mathcal{Z} = \{z_1, z_2, \ldots, z_N\} \in \mathbb{R}^d$. The key idea is that the high-dimensional data $x$ can be explained by the substantially lower-dimensional and semantically meaningful latent representation $z$, which is mapped back to the higher-dimensional space of observations $x$. $\mathcal{F}$ is to be learned under the data condition $x$ (i.e., $p(f^c | x)$), which can separate different factors of variation in the data, making downstream tasks easier, such as RFFI. In DRL, $p(f^c | x)$ is often approximated by $p(f^c | z)$. This will be discussed in Section \ref{subsec:DRL}.

\subsection{Disentanglement Representation Learning}\label{subsec:DRL}

A DRL process makes latent unit ideally captures a single explanatory factor, while being relatively invariant to changes in other factors \cite{Bengio_Representation_2014}. In such representations, perturbing a single latent dimension should result in changes in only one factor of the data while keeping other factors invariant. This characteristic is believed to enhance robustness and interpretability in downstream tasks, as it more closely aligns with physical process the generating observed data.

Inspired by the above theory, the real-world signal is generated by a twostep process \cite{Locatello_Challenging_2019}: 
\begin{enumerate}
    \item Sampling latent representations $z \in \mathbb{R}^{N}$ from distribution $p(z)$. $z$ corresponds to semantically meaningful conditional factors $f^c$ of the observations.
    \item Generating observed data $x$ through conditional distribution $p(x|z)$. 
\end{enumerate}

Using the RF signal as an example, the factors include RFF, modulation, channel condition, and receiver, etc. Fig. \ref{cross_space_2} illustrates three factors (RFF, modulation, and receiver). The conditional distributions of the factors, denoted as $\{p(x | f^c) | c \in \mathcal{C} = \{\text{RFF, modulation, receiver}\}\}$, can be individually represented as curved surfaces in Fig. \ref{cross_space}. The relationship between the conditional distribution $p(x | f^c)$ and the data distribution $p(x)$ can be expressed as
\begin{equation}
  \begin{aligned}
    p(x) &= \int p(x|f^c)p(f^c)df^c \\&= \int\cdots\int p(x|f^1,\ldots,f^N)p(f^1)\ldots p(f^N)df^1\ldots df^N
  \end{aligned}
\end{equation} 

This relationship holds since $f^1, \ldots, f^N$ and $x$ form a V-structure in the Probabilistic Graphical Model (PGM) shown in Fig. \ref{PGM}. The score function of the conditional distribution can be expressed using Bayes' rule as
\begin{equation}
  \begin{aligned}
    \nabla_{x} \log p(x | f^c) = \nabla_{x} \log p(f^c | x) + \nabla_{x} \log p(x)
  \end{aligned}\label{eq:bayes}
\end{equation}

Data can be sampled conditioned on the factor $f^c$ by using the score function of $p(x | f^c)$, as shown in Fig. \ref{cross_space}. One objective of DRL is to model $\nabla_{x} \log p(x | f^c)$ for each factor $f^c$, which represents how the factor $f^c$ influences the data distribution $p(x)$. According to Eq. \eqref{eq:bayes}, $\nabla_{x} \log p(f^c | x)$ can be learned instead, which corresponds to the arrows pointing towards the curved surface in Fig. \ref{cross_space}.

However, directly mapping data to $f^c$ often suffers from the problem of unidentifiability: even if there are multiple reasonable factorization schemes, it is difficult for the model to determine which one is the true generative process unless sufficient inductive bias is introduced \cite{Locatello_Challenging_2019}.

In fact, a set of representations $\mathcal{Z} = \{z^c | c=1, \ldots, N\}$ can be obtained to approximate the factors $\mathcal{F}$. As illustrated in Fig. \ref{fig:exp_mod_com}, three main requirements must be satisfied for the representations $\mathcal{Z}$ to be considered a good approximation of the factors $\mathcal{F}$ \cite{Eastwood_Framework_2018, carbonneau_Measuring_2022, wang_Disentangled_2024}. Utilizing these representations $\mathcal{Z}$, the gradient $\nabla_{x} \log p(z^c | x)$ can serve as an approximation for $\nabla_{x} \log p(f^c | x)$.
\begin{enumerate}
  \item \textit{Explicitness}: There should be a complete bijection, $\mathcal{Z} \mapsto \mathcal{F}$. In other words, it should be possible to retrieve any factor $f^c$ from a point in the representation space $\mathcal{Z}$.
  \item \textit{Modularity}: $f^{c_1}$ should be informative to predict $z^{c_1}$, and $f^{c_1}$ should not contain information about $z^{c_2}$ for $c_1 \neq c_2$.
  \item \textit{Compactness}: The influence of $f^{c_1}$ on $z^{c_1}$ should be minimized (however, enforcing \textit{compactness} could be counterproductive in practical scenarios \cite{Ridgeway_learning_2018}).
\end{enumerate}

Among the three requirements, \textit{Explicitness} is a basic necessity, and without a certain degree of \textit{Explicitness}, it is impossible to reasonably compare factor-representation relationships. However, the value of \textit{Explicitness} does not directly reflect the degree of \textit{Modularity} or \textit{Compactness}. \textit{Modularity} is the most critical metric that directly reflects whether each factor is confined to a specific subset of the representation space. Although \textit{Compactness} describes how representations concentrate on a factor, a factor often requires multi-dimensional representation in practical problems, which makes it the least important.

\subsection{Diffusion Probablistic Model for Representation}\label{subsec:DPM}

According to Eq. \eqref{eq:bayes}, constructing $\log p(x | f^c)$ in DRL requires $\log p(f^c | x)$ and $p(x)$. In the proposed method, we use the generative model DPM to construct $\log p(x | f^c)$ .

As the originator of DRL, the principle of Variational Auto-Encoder (VAE) explains how to achieve disentanglement through a generative model. A key feature of these methods is the dimension-wise structure of the disentangled representations, where different dimensions of the latent vector correspond to distinct factors. VAE-based methods model data distributions through maximum likelihood estimation by optimizing $p(x)$
\begin{equation}
  \begin{aligned}
    \log p({x})=D_{KL}\left(q_\phi({z}|{x})\|p({z}|{x})\right)+\mathcal{L}(\theta,\phi;{x},{z})
  \end{aligned}\label{eq:VAE}
\end{equation}
The first term of Eq. \eqref{eq:VAE} represents the Kullback-Leibler (KL) divergence between the variational posterior distribution $q_\phi({z}|{x})$ and the true posterior distribution $p({z}|{x})$. The second term is referred to as the Evidence Lower Bound (ELBO), under the condition that the KL divergence term is non-negative. By maximizing the ELBO, a tight lower bound is provided for $p(x)$
\begin{equation}
  \begin{aligned}
    \mathcal{L}(\theta,\phi;{x},{z})=&-D_{KL}\left(q_\phi({z}|{x})\|p({z})\right)\\& +\mathbb{E}_{q_\phi({z}|{x})}\left[\log p({x}|{z})\right]
  \end{aligned}\label{eq:ELBO}
\end{equation}
where $\mathbb{E}_{q_\phi({z}|{x})}\left[\log p({x}|{z})\right]$ is responsible for the reconstruction, while the KL divergence quantifies the discrepancy between the variational posterior distribution $q_\phi({z}|{x})$ and the prior distribution $p(z)$. Typically, a \textbf{Gaussian distribution} is selected for $p(z)$, thereby ensuring that the KL term enforces independent constraints on the representations.

The DPM is widely recognized as a better generator, which also operates through Gaussian distributions. A common DPM learns a model $\epsilon(x_t,t)$ to predict the noise added to a sample $x_t$, where $x_t$ is the $t$-th processed noised sample of the original sample $x_0$. In the forward process, noise is gradually added to the data until it becomes pure noise. At the $t$-th step, the data ${x}_t$ is obtained by adding noise to the previous step's data ${x}_{t-1}$
\begin{equation}
  \begin{aligned}
    q({x}_t | {x}_{t-1}) &= \mathcal{N}({x}_t; \sqrt{1 - \beta_t} \cdot {x}_{t-1}, \beta_t \cdot \mathbf{I})\\
    q({x}_{1:T}|{x}_0) &= \prod_{t=1}^T q({x}_t|{x}_{t-1})
  \end{aligned}
\end{equation}
where $\beta_t$ represents the noise variance added at the $t$-th step, and $\mathcal{N}(\mu, \sigma^2)$ is a Gaussian distribution with mean $\mu$ and variance $\sigma^2$. Defining $\alpha_t = 1 - \beta_t$ and $\bar{\alpha}_t = \prod_{s=1}^t \alpha_s$, $x_t$ can be sampled with the formula $x_t \sim \mathcal{N}(x_t; \sqrt{\bar{\alpha}_t} x_0, (1-\bar{\alpha}_t))$, i.e., $x_t = \sqrt{\bar{\alpha}_t} x_0 + \sqrt{1-\bar{\alpha}_t} \epsilon$.

Over multiple iterations, the data is progressively corrupted by noise and eventually approaches a standard normal distribution. The reverse process learns how to gradually remove noise from the noisy data $x_T$ and recover the original data $x_0$, where $T$ refers to the maximum iteration. Let the denoising function be parameterized as $p({x}_{t-1} | {x}_t)$, which is typically modeled as
\begin{equation}
  \begin{aligned}
    p({x}_{t-1} | {x}_t) &= \mathcal{N}({x}_{t-1}; \mu({x}_t, t), \Sigma({x}_t, t))\\
    p({x}_{0:T}) &= p({x}_T)\prod_{t=1}^T p({x}_{t-1}|{x}_t)
  \end{aligned}
\end{equation}
where $\mu({x}_t, t)$ and $\Sigma({x}_t, t)$ are the mean and variance predicted by a neural network.

During the forward process of the DPM, the data is gradually mapped to a simple Gaussian distribution while preserving global information. No factorization assumption is required during generation, allowing the model to freely sample instances that conform to all statistical properties of the original data distribution $p(x)$. If $f^c$ is introduced into the basis of DPM as $p(f^c|x)$, it will adapt to the disentangling task by modeling the data distribution $p(x|f^c)$ in Eq. \eqref{eq:bayes} \cite{yang_DisDiff_2023,wu_Factorized_2024}. The introduction of $p(f^c|x)$ information in DPM will be detailed in Section \ref{subsec:structure}.

\section{Methodology}\label{sec:model}

\begin{figure*}[!t]
  \centering
  {\includegraphics[width=2\columnwidth]{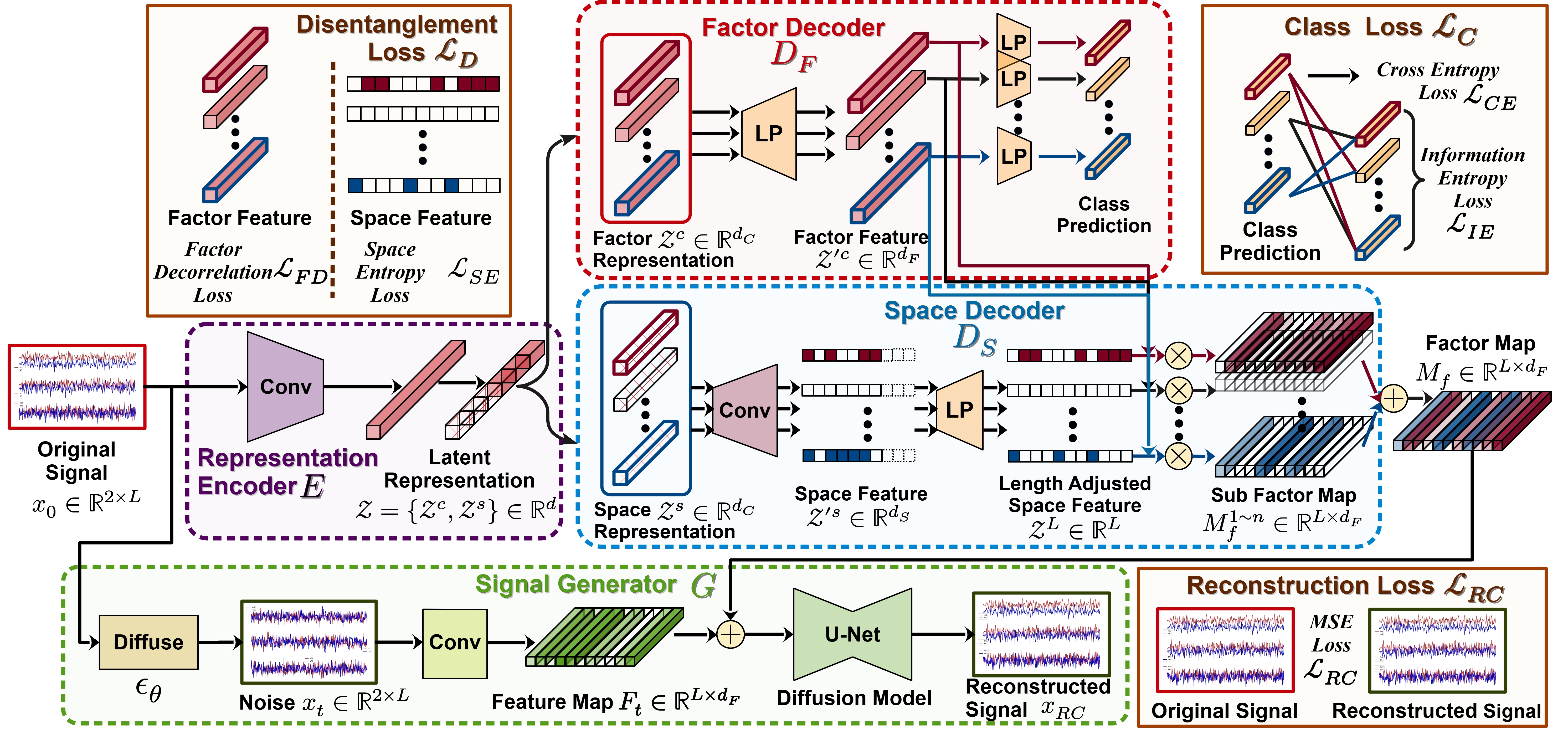}\hspace{5mm}
  \caption{The proposed model framework. The raw signal $x_0$ is processed by the representation encoder $E$, yielding representations $\mathcal{Z}$ for both factor $\mathcal{Z}^c$ and corresponding space information $\mathcal{Z}^s$. These two representations $\mathcal{Z}^c$ and $\mathcal{Z}^s$ are separately decoded and then combined by element-wise multiplication followed by summation to generate the factor map $M_f$. The factor map serves as the conditional input to the signal generator $G$, governing the generation of the reconstructed signal $p(x|z^c)$.}\label{fig:framework}}
\end{figure*}

\subsection{System Overview}\label{subsec:structure}

Given an input $x_0 \in \mathcal{X}$, the objective of DRL is to simultaneously learn the representation $z^c$ for each factor $f^c \in \mathcal{F}$ and its associated gradient field $\nabla_{x_t} \log p(z^c | x)$. To meet the requirements of \textit{explicitness}, \textit{modularity}, and \textit{compactness} in Section \ref{subsec:DRL}, the proposed framework is designed in Fig. \ref{fig:framework}. The Representation Encoder $E$ extracts the representation corresponding to each factor. The Factor Decoder $D_F$ and Space Decoder $D_S$ further process the representations, incorporating a Disentanglement Loss $\mathcal{L}_{D}$ to enforce the aforementioned principles of \textit{explicitness}, \textit{modularity}, and \textit{compactness}. Classifiers placed after $D_F$ performs the classification task, while the Signal Generator $G$ enables conditional signal generation.

\subsubsection{Representation Encoder $E$}
Specifically, an encoder $E$ is employed to obtain the representations for each factor representation $z^c$ as
\begin{equation}
  \begin{aligned}
    E(x_0) &= \mathcal{Z} = \{\mathcal{Z}^c, \mathcal{Z}^s\} \\
           &= \{\underbrace{{z}^{c_1},{z}^{c_2},\dots,{z}^{c_N}}_{N\text{ factor representations}}, \underbrace{{z}^{s_1},{z}^{s_2},\dots,{z}^{s_N}}_{N\text{ space representations}}\}
  \end{aligned}\label{eq:encoder}
\end{equation}
where $\mathcal{Z} \in \mathbb{R}^d$ is the concatenation of factor representations ${z}^{c_1},{z}^{c_2},\dots,{z}^{c_N} \in \mathbb{R}^{d_C}$ and space representations ${z}^{s_1},{z}^{s_2},\dots,{z}^{s_N} \in \mathbb{R}^{d_C}$. Each $z^c$ corresponds to a $z^s$, where $z^s$ denotes the position of $z^c$ in the representation space.

\subsubsection{Factor Decoder $D_F$ and Space Decoder $D_S$}

As shown in Fig. \ref{fig:framework}, the Factor Decoder $D_F$ expands the factor representation $z^c$ provided by $E$ from $d_C$ to $d_F$, producing a Factor Feature. The Space Decoder $D_S$ explicitly expands the dimension of $d_C$ corresponding to the space representation space to $d_S$, producing a Space Feature. The design idea behind the two modules is as follows. 

For the Factor Decoder $D_S$, it meets the following principle by continuing to process the representation $z^c$. By limiting the dimension of $d_C$ to match the number of factors, $z^c$ can fully represent all factors and meet the requirement of \textit{explicitness}. Additionally, by minimizing the correlation between each $z^c$, the \textit{modularity} of $z^c$ can be improved. 

For the Space Decoder $D_S$, the overlap of the space occupied by each $z^c$ can be observed. \textit{Compactness} aims for the representation to occupy the smallest possible space, so reducing the overlap of space occupation by each $z^c$ helps minimize the representation space occupancy. To achieve these goals, loss functions are designed in Section \ref{subsec:loss}. The Space Decoder is expected to explicitly extract space information in $S^n \, (n = 1, 2, \dots, N)$ via an interpretable masked feature. A CNN with upsampling modules is adopted as the space decoder $D_S$. The space feature $s^n$ is decoded by
\begin{equation}
  \begin{aligned}
    \mathcal{Z'}^{s_n} &= \mathrm{Softmax}(D_S(z^{s_1}), D_S(z^{s_2}), \dots, D_S(z^{s_N}))_n
  \end{aligned}\label{eq:space_decoder}
\end{equation}
where the Softmax function normalizes the $N$ space values, ensuring that each value in the space feature is a probability between 0 and 1.

\subsubsection{Factor Map $M_f$ and Signal Generator $G$}

To represent the $n$-th factor concept interpretably, the factor representation $z^{c_n}$ and the space representation $z^{s_n}$ need to be aggregated to form a conditional input $f^c$ that conforms to the data distribution $p(x|f^c)$ in Eq. \eqref{eq:bayes}.

First, $z^{c_n} \in \mathbb{R}^{d_C}$ is mapped to $z'^{c_n} \in \mathbb{R}^{d_F}$ through linear projection (LP) to adjust the feature dimension to $d_F$. Next, the space representation $z^{s_n}$ is passed through a convolutional layer and a linear projection to obtain the result $z_L$ of size ${R}^L$. Then, $z_L$ is multiplied with $z'^{c_n}$ to obtain the conditional factor map $M^n_f \in \mathbb{R}^{L \times d_F}$, where $L$ represents the length of the original signal, and $d_F$ is the dimension matching the feature map $F_t$ of the DPM signal generator $G$.

Therefore, $F_t$ extracts normally distributed noise through the diffusion process to obtain the unconditional features of the signal, while $M_f$ obtains the conditional features of the signal through $z^c$ and $z^s$, which carry semantic information. This allows $G$ to model the data distribution $p(x|f^c)$. The reconstructed signal $x_{RC}$, reflecting $p(x|f^c)$, generated by the generator $G$ with the factor map $M_f$, can be expressed as
\begin{equation}
  \begin{aligned}
    M_{f}^n &= \mathrm{LP}({z'}^{s_n}) \cdot {z'}^{c_n} \\
    x_{RC} &= G(F_t, M_f)
  \end{aligned}\label{eq:factor_map}
\end{equation}

\subsection{Loss Function Design}\label{subsec:loss}

Inspired by \cite{wang_Disentangled_2024, wu_Factorized_2024, yang_DisDiff_2023}, the loss function for model training consists of three components: disentanglement loss $\mathcal{L}_{D}$, reconstruction loss $\mathcal{L}_{RC}$, and classification loss $\mathcal{L}_{C}$.

Among these, the reconstruction loss $\mathcal{L}_{RC}$ ensures that the representations correspond to the factors while recover the original data according to $p(x|f^c)$. The classification loss $\mathcal{L}_{C}$, involves classification tasks targeting different factors. The labels of these factors provide significant inductive bias \cite{Locatello_Challenging_2019}, which aids the disentangling process. Disentanglement loss $\mathcal{L}_{D}$ defines the goal of disentangling, ensuring that the representations $\mathcal{Z}$ meet the requirements of \textit{explicitness}, \textit{modularity}, and \textit{compactness} \cite{carbonneau_Measuring_2022}. The total loss function is defined as
\begin{equation}
  \begin{aligned}
    \mathcal{L} = \lambda_{RC} \mathcal{L}_{RC} + \lambda_{D} \mathcal{L}_{D} + \lambda_{T} \mathcal{L}_{C}
  \end{aligned}\label{eq:total_loss}
\end{equation}
where $\lambda_{RC}$, $\lambda_{D}$, and $\lambda_{T}$ are the trade-off parameters.

\subsubsection{Disentanglement Loss $\mathcal{L}_{D}$}

In the factor representation space $\mathcal{Z}^{c}$, it is expected that $z^{c_1}$ and $z^{c_2}$ from different dimensions are decorrelated for $c_1 \neq c_2$ to achieve \textit{modularity}. Therefore, the factor decorrelation loss $\mathcal{L}_{FD}$ is introduced by minimizing the covariance $\mathrm{cov}(z^{c_1}, z^{c_2})$ as
\begin{equation}
  \begin{aligned}
    \min_{E}\mathcal{L}_{FD} &= \sum_{c_1 \neq c_2} \left|\mathrm{cov}(z^{c_1}, z^{c_2})\right| \\
    &= \sum_{c_1 \neq c_2} \left|\frac{1}{N-1} \sum_{m=1}^{M} \Vert z_m^{c_1} \Vert_2^\top \Vert z_m^{c_2} \Vert_2\right|
  \end{aligned}
\end{equation}
where $z^{c_1}$ and $z^{c_2}$ are the representations of different factors, and $M$ is the number of training samples. The number of factor representations $c_n$ is artificially set to match the number of factors, ensuring that the feature space composed of all factors satisfies the \textit{explicitness} requirement. Each factor occupies a non-overlapping region in the space, ensuring that the factors remain uncorrelated, thus satisfying \textit{modularity}.

In the space representation $\mathcal{Z}^{s_n}$, the $i$-th element represents the probability that the information corresponding to the $n$-th factor representation $\mathcal{Z}^{c_n}$ appears at the $i$-th position in the representation space. It is desirable for $z^{s_{n1}}_i$ to be close to 1, while the elements of $\{ z^{s_{n2}}_i \mid n_1 \neq n_2 \}$ should be close to 0. This condition ensures minimal uncertainty when selecting the factor for this particular representation. To achieve this, the space entropy loss is introduced as
\begin{equation}
  \begin{aligned}
    \min_{E, D_S} \mathcal{L}_{SE} &= \frac{1}{d_S \times M} \sum_{i,m} \frac{1}{N} \sum_{n=1}^{N} - z^{s_n}_{i,m} \log(z^{s_n}_{i,m})
  \end{aligned}
\end{equation}
where $d_S$ is the dimension of the space feature, $N$ is the number of factors, $m$ denotes the $m$-th signal sample among the $M$ total samples, and $z^{s_n}_{i,m}$ represents the $i$-th value of the space position corresponding to $z^{c_n}_m$. This operation minimizes the occupancy of each factor in the space, while limiting the dimensional size of the spatial feature, aligning with the \textit{compactness} assumption.

Together, $\mathcal{L}_{FD}$ and $\mathcal{L}_{SE}$ apply constraints to the factor representation $\mathcal{Z}^c$ and space representation $\mathcal{Z}^s$. They work to minimize the correlations among all $\mathcal{Z}^c$ while enhancing the distinguishability of different factors within the feature space using $\mathcal{Z}^s$. The total disentanglement loss $\mathcal{L}_{D}$ is then given by
\begin{equation}
  \begin{aligned}
    \mathcal{L}_{D} = \lambda_{FD} \mathcal{L}_{FD} + \lambda_{SE} \mathcal{L}_{SE}
  \end{aligned}\label{eq:D}
\end{equation}
where $\lambda_{FD}$ and $\lambda_{SE}$ are the weights for the factor decorrelation and space entropy losses, respectively.

\subsubsection{Reconstruction Loss $\mathcal{L}_{RC}$}

As discussed in Section \ref{subsec:DPM}, the reconstruction loss between the input signal and the signal reconstructed by the DPM can be formulated as
\begin{equation}
  \begin{aligned}
    \min_{E,D_F,D_S,G} \mathcal{L}_{RC} = \mathbb{E}_{x,t} \| G(x_t, M_f, t) - x \|_2^2
  \end{aligned}\label{eq:RC}
\end{equation}
where $x \sim p(x)$ denotes the input signal. As shown in Fig. \ref{fig:framework}, at time step $t$ of the DPM forward process, the noised signal $x_t = x_0 + n_t$ is produced by the diffusion process, where $n_t \sim \mathcal{N}(0, \sigma_t^2 I)$. Here, $\sigma_t$ represents the noise schedule at step $t$, and $\sigma_t^2 = \beta_t$ \cite{ho_Denoising_2020}. The reconstructed signal $x_{RC}$ is generated at time $t$ by the generator $G(x_t, M_f, t)$, which incorporates the factor map $M_f$.

\subsubsection{Class Loss $\mathcal{L}_{C}$}

For each factor, the labels inside the factor are introduced as inductive biases to guide the disentangling process. The Cross-Entropy (CE) loss is introduced to quantify the classification error $\mathcal{L}_{CE}$ as
\begin{equation}
  \begin{aligned}
    \min_{E, D_F} \mathcal{L}_{CE} &= - \sum_{n=1}^{N} \sum_{k=1}^{K} y^n_k \log \hat{y}^n_k
  \end{aligned}\label{eq:CE}
\end{equation}

Next, the Information Entropy (IE) loss $\mathcal{L}_{IE}$ is introduced to further leverage classification tasks across different factors, enhancing the level of disentangling. For the features of the $n_1$-th factor among the $N$ factors, in addition to being inputted into the $n_1$-th classifier to calculate the classification loss $\mathcal{L}_{CE}$, these features are also inputted into all other $n_2$ classifiers ($n_1 \neq n_2$), yielding $N-1$ classification results $\hat{y}^{n_1}_{n_2}$. The entropy loss is then computed, and the summation of these losses gives the complete $\mathcal{L}_{IE}$, defined as
\begin{equation}
  \begin{aligned}
    \max_{E, D_F} \mathcal{L}_{IE} &= - \sum_{n_1=1}^{N} \mathcal{L}^{f^{{n_1}} \rightarrow f^{{n_2}}}_{IE}, \quad n_1 \neq n_2 \\
    &= - \sum_{n_1=1}^{N} \sum_{n_2=1}^{N} \hat{y}^{n_1}_{n_2} \log \hat{y}^{n_1}_{n_2}, \quad n_1 \neq n_2
  \end{aligned}
\end{equation}

Unlike the previous losses, the Information Entropy (IE) loss $\mathcal{L}_{IE}$ is maximized to encourage disentangling. As an example, consider the modulation scheme and RFF factors. A larger value of $\mathcal{L}^{modulation \rightarrow RFF}_{IE}$ indicates that when the RFF classifier predicts the modulation scheme’s factor feature, the output vector $\hat{y}^{modulation}_{RFF}$ tends to exhibit a more uniform distribution. This makes it difficult for the RFF classifier to predict the RFF label based on the modulation scheme’s factor feature. From an information theory perspective, this means that the information about the modulation scheme and RFF is independent within the factor feature, preventing the RFF classifier from extracting RFF-related information, thus achieving the goal of disentangling.

The class loss $\mathcal{L}_{C}$ is formulated with trade-off parameters $\lambda_{CE}$ and $\lambda_{IE}$ to balance the above two losses. The overall loss function is
\begin{equation}
  \begin{aligned}
    \mathcal{L}_{C} = \lambda_{CE} \mathcal{L}_{CE} + \lambda_{IE} \mathcal{L}_{IE}
  \end{aligned}\label{eq:T}
\end{equation}


\section{Experimental Setup}\label{sec:exp_setup}

Experiments are conducted on three datasets to validate the effectiveness of the proposed method. These include two public RFF datasets: \textbf{POWDER} \cite{ReusMuns_Trust_2020} and \textbf{WiSig-ManySig} \cite{Hanna_WiSig_2022}, as well as a self-collected dataset based on the public ideal dataset \textbf{TorchSig-Real} \cite{Boegner_Large_2022}. A summary of the key parameters for each dataset is presented in Table \ref{tab:dataset}, with each contains 3 factors as $f^1$, $f^2$, $f^3$. All models are implemented on an NVIDIA GeForce RTX 2080 platform and trained using PyTorch 2.0.1 and Python 3.10.14.

\begin{table}[!t]
  \centering
  \renewcommand{\arraystretch}{1}
  \tabcolsep=.12cm
  \caption{Dataset parameters}
    \begin{tabular}{ccccc}
    \toprule
    \textbf{Dataset} & \textbf{Factor} & \textbf{Factor No.} & \textbf{Sig. No.} & \textbf{Sig. Size} \\
    \midrule
    \multirow{3}[0]{*}{POWDER} & Day(Channel) & 2     & \multirow{3}[0]{*}{2000} & \multirow{3}[0]{*}{512} \\
          & Standard(Protocal) & 3     &       &  \\
          & Transmitter(RFF) & 4     &       &  \\
    \midrule
    \multirow{3}[0]{*}{WiSig-ManySig} & Day(Channel) & 4     & \multirow{3}[0]{*}{1000} & \multirow{3}[0]{*}{256} \\
          & Receiver & 12    &       &  \\
          & Transmitter(RFF) & 6     &       &  \\
    \midrule
    \multirow{3}[0]{*}{TorchSig-Real} & SNR(Channel) & 7     & \multirow{3}[0]{*}{2000} & \multirow{3}[0]{*}{512} \\
          & Modulation & 5     &       &  \\
          & Transmitter(RFF) & 7     &       &  \\
    \midrule
    \end{tabular}\label{tab:dataset}%
\end{table}%

\begin{table*}[!t]
  \centering
  \renewcommand{\arraystretch}{1}
  \tabcolsep=.1cm
  \caption{The DRL performance reflected by each metric and the characteristics of each metric}
    \begin{threeparttable}
    \begin{tabular}{l ccc ccc}
    \toprule

    \multirow{2}{*}{\thead{\\ \\ \textbf{Metric}}}
    &\multicolumn{3}{c}{\textbf{Measurable Performace}}&\multicolumn{3}{c}{\textbf{Characteristic}}
    \cr
    \cmidrule(lr){2-4} \cmidrule(lr){5-7}
    & \textbf{Explicitness} & \textbf{Modularity} & \textbf{Compactness} & \thead{\textbf{Noise} \\ \textbf{Robustness}} & \thead{\textbf{Non-Measured Factor} \\ \textbf{Robustness}} & \thead{\textbf{Hyperparameter} \\ \textbf{Requirement}} 
    \cr

    \midrule
    \tnote{1} $\mathrm{Z_{\text{diff}}^{+}}$ \cite{higgins_beta_2017} & \textcolor{lightgray}{\ding{55}} & \textcolor{lightgray}{\ding{55}} & \textcolor{lightgray}{\ding{55}} & \ding{51} & \ding{51} & \ding{51}\\
    $\mathrm{Z_{\text{min}}^{+}}$ \cite{kim_disentangling_2019} & \textcolor{lightgray}{\ding{55}} & \textcolor{lightgray}{\ding{55}} & \textcolor{lightgray}{\ding{55}} & \ding{51} & \ding{51} & \ding{51}\\
    $\mathrm{Z_{\text{max}}^{+}}$ \cite{kim_relevance_2019} & \textcolor{lightgray}{\ding{55}} & \textcolor{lightgray}{\ding{55}} & \textcolor{lightgray}{\ding{55}} & \textcolor{lightgray}{\ding{55}} & \textcolor{lightgray}{\ding{55}} & \ding{51}\\

    \midrule
    $\mathrm{Modularity \enspace Score^{+}}$ \cite{Ridgeway_learning_2018} & \textcolor{lightgray}{\ding{55}} & \ding{51} & \textcolor{lightgray}{\ding{55}} & \ding{51} & \textcolor{lightgray}{\ding{55}} & \textcolor{lightgray}{\ding{55}}\\
    $\mathrm{IRS^{+}}$ \cite{Suter_Robusty_2019} & \ding{51} & \ding{51} & \textcolor{lightgray}{\ding{55}} & \textcolor{lightgray}{\ding{55}} & \textcolor{lightgray}{\ding{55}} & \textcolor{lightgray}{\ding{55}}\\
    $\mathrm{DCIMIG^{+}}$ \cite{Sepliarskaia_How_2021} & \ding{51} & \ding{51} & \textcolor{lightgray}{\ding{55}} & \textcolor{lightgray}{\ding{55}} & \ding{51} & \textcolor{lightgray}{\ding{55}}\\
    $\mathrm{JEMMIG^{-}}$ \cite{Do_Theory_2021} & \ding{51} & \ding{51} & \ding{51} & \textcolor{lightgray}{\ding{55}} & \ding{51} & \textcolor{lightgray}{\ding{55}}\\
    $\mathrm{SAP^{+}}$ \cite{Kumar_variational_2018} & \ding{51} & \textcolor{lightgray}{\ding{55}} & \ding{51} & \textcolor{lightgray}{\ding{55}} & \ding{51} & \textcolor{lightgray}{\ding{55}}\\
    $\mathrm{Explicitness \enspace Score^{+}}$ \cite{Ridgeway_learning_2018} & \ding{51} & \textcolor{lightgray}{\ding{55}} & \textcolor{lightgray}{\ding{55}} & \textcolor{lightgray}{\ding{55}} & \ding{51} & \textcolor{lightgray}{\ding{55}}\\

    \bottomrule
    \end{tabular}%
    \begin{tablenotes}
      \footnotesize 
      \item[1] The superscript `+' indicates that a higher value corresponds to better performance, while `-' indicates the opposite. 
    \end{tablenotes}
  \end{threeparttable}%
  \label{tab:metrics_properties}%
\end{table*}%

\subsection{Dataset1: POWDER}

The POWDER dataset was acquired using USRP X310 SDRs and USRP B210 SDRs as Base Stations (BSs). A fixed USRP B210 receiver authenticates transmissions from four surrounding BSs, each transmitting standards-compliant Wi-Fi (IEEE 802.11ac), 4G LTE, and 5G NR frames generated through MATLAB toolboxes. I/Q samples were collected at a rate of 5~MS/s for Wi-Fi and 7.68~MS/s for LTE and 5G NR, all centered at a frequency of 2.685~GHz. The distances between the BSs and the central receiver ranged from 300~m to 1~km. For each BS, I/Q samples were recorded during sessions separated by 10 seconds within a day, and the entire process was repeated across 2 different days.

\subsection{Dataset2: WiSig-ManySig}
WiSig is a Wi-Fi signal dataset. In the original one-day capture process, signals from each transmitter were recorded sequentially. Each transmitter was configured to transmit random byte streams to a Wi-Fi access point over the 2.4~GHz band, using identical spoofed MAC and IP addresses. During each transmission, all USRP receivers were configured to capture signals over the same bandwidth for approximately 0.5~s. WiSig-ManySig is a subset of the WiSig dataset, consisting of signals captured from 6 transmitters and 12 receivers over 4 distinct days.

\subsection{Dataset3: TorchSig-Real}

TorchSig is an public simulated signal dataset that includes a variety of unique signal modulations. In our experiments, we selected 25 modulation schemes from TorchSig, named it as TorchSig-Real, and categorized it into 5 major groups: ASK, PSK, PAM, QAM, and QAM\_cross. Within a signal-to-noise ratio (SNR) range of \(-15~\mathrm{dB}\) to \(15~\mathrm{dB}\) at 5~dB intervals, these communication signals were transmitted using 7 HackRF-One transmitters via feeder lines to a USRP B200 receiver. The RF frequency was set to 330~MHz, the carrier frequency to 1~MHz, the sampling rate to 16~MS/s, and the symbol duration was fixed at 20\(\mu\)s for all transmissions.

\section{Experimental Results}\label{sec:exp}

In Section~\ref{subsec:comparison}, we first compare classical disentangled representation learning (DRL) methods to validate our model’s disentangling performance from the perspectives of \textit{explicitness}, \textit{modularity}, and \textit{compactness}. We then demonstrate the effectiveness of the proposed module designs through ablation studies. Subsequently, Sections~\ref{subsec:classification} and~\ref{subsec:generalization} evaluate the performance of DRL on classification tasks and conditional signal generation.

\subsection{Metrics Comparison and Analysis}\label{subsec:comparison}

\subsubsection{Metrics Properties}\label{subsubsec:metrics_properties}
We utilized $Z_{\text{diff}}$ \cite{higgins_beta_2017}, $Z_{\text{min}}$ \cite{kim_disentangling_2019}, $Z_{\text{max}}$ \cite{kim_relevance_2019}, Modularity Score \cite{Ridgeway_learning_2018}, Interventional Robustness Score (IRS) \cite{Suter_Robusty_2019}, DCIMIG \cite{Sepliarskaia_How_2021}, JEMMIG \cite{Do_Theory_2021}, Attribute Predictability Score (SAP) \cite{Kumar_variational_2018}, and Explicitness Score \cite{Ridgeway_learning_2018} for comparison experiment of representation. The measurable performance and characteristic of each metric are shown in Table \ref{tab:metrics_properties}. Each metric reflects one or more disentangling aspects of \textit{explicitness}, \textit{modularity}, and \textit{compactness}.

Each metric has mainly three characteristics. \textbf{Noise robustness} indicates whether introducing noise into a perfect representation affects the metric value. \textbf{Non-Measured Factor Robustness} reflects whether the metric varies if the representation captures only part of a factor rather than the entire factor. \textbf{Hyperparameter requirement} indicates whether the metric's measurement depends on hyperparameters, which may impact the score.

\subsubsection{Comparison Results}

\begin{table*}[!t]
  \centering
  \renewcommand{\arraystretch}{1}
  \tabcolsep=.15cm
  \caption{Comparison on Metrics that Quantify Explicitness, Modularity, and Compactness}
    \begin{tabular}{clrrrrrrrrr}
    \toprule
    Dataset & \multicolumn{1}{c}{Method} & \multicolumn{1}{c}{$Z_{\text{diff}}$} & \multicolumn{1}{c}{$Z_{\text{min}}$} & \multicolumn{1}{c}{$Z_{\text{max}}$} & \multicolumn{1}{c}{Modularity} & \multicolumn{1}{c}{IRS} & \multicolumn{1}{c}{DCIMIG} & \multicolumn{1}{c}{JEMMIG} & \multicolumn{1}{c}{SAP} & \multicolumn{1}{c}{Explicitness} \\
    \midrule
    \multirow{6}[0]{*}{POWDER} & VAE   & 3.319E-01 & 3.713E-01 & 3.750E-01 & 9.875E-01 & 4.863E-01 & 3.438E-02 & 3.758E-01 & 1.700E-04 & 6.615E-02 \\
          & $\beta$-$\mathrm{VAE}_H$ & 3.319E-01 & 3.663E-01 & 4.062E-01 & 9.684E-01 & 4.693E-01 & 2.593E-02 & 3.826E-01 & 8.800E-05 & 4.955E-02 \\
          & $\beta$-$\mathrm{VAE}_B$ & 3.319E-01 & 4.175E-01 & 3.944E-01 & 9.981E-01 & 2.552E-01 & 4.089E-02 & 3.822E-01 & 1.325E-04 & \underline{7.004E-02} \\
          & FactorVAE & 3.312E-01 & \underline{5.725E-01} & 4.487E-01 & \underline{9.991E-01} & 4.738E-01 & 1.789E-02 & 6.476E-01 & 2.445E-04 & 5.275E-02 \\
          & $\beta$-TCVAE & 3.251E-01 & \textbf{6.134E-01} & \underline{4.695E-01} & \textbf{9.984E-01} & \underline{5.083E-01} & \underline{4.795E-02} & \underline{3.510E-01} & \underline{1.881E-03} & 5.571E-02 \\
          & \textbf{Proposed} & \textbf{1.000E+00} & 3.360E-01 & \textbf{8.718E-01} & 9.241E-01 & \textbf{5.745E-01} & \textbf{7.528E-01} & \textbf{3.382E-01} & \textbf{3.704E-02} & \textbf{9.994E-01} \\
    \midrule
    \multirow{6}[0]{*}{\thead{WiSig- \\ ManySig}} & VAE   & 1.000E+00 & 6.862E-01 & 6.675E-01 & 7.682E-01 & 4.286E-01 & 1.364E-01 & \underline{2.667E-01} & 7.134E-03 & 6.829E-01 \\
          & $\beta$-$\mathrm{VAE}_H$ & 1.000E+00 & 6.631E-01 & \underline{7.069E-01} & 8.048E-01 & \underline{5.078E-01} & 1.470E-01 & 3.810E-01 & \underline{7.124E-02} & 6.632E-01 \\
          & $\beta$-$\mathrm{VAE}_B$ & 1.000E+00 & 8.225E-01 & 6.994E-01 & 8.152E-01 & 3.786E-01 & 1.079E-01 & \textbf{2.497E-01} & 2.110E-02 & 6.701E-01 \\
          & FactorVAE & 1.000E+00 & \textbf{8.744E-01} & 6.562E-01 & \underline{8.163E-01} & 4.595E-01 & 1.370E-01 & 3.063E-01 & 4.543E-02 & 6.900E-01 \\
          & $\beta$-TCVAE & 1.000E+00 & \underline{8.588E-01} & 6.787E-01 & 7.942E-01 & 4.440E-01 & \underline{1.788E-01} & 2.687E-01 & 4.987E-02 & \underline{6.928E-01} \\
          & \textbf{Proposed} & \textbf{1.000E+00} & 3.270E-01 & \textbf{9.815E-01} & \textbf{9.764E-01} & \textbf{5.614E-01} & \textbf{4.854E-01} & 2.946E-01 & \textbf{7.849E-02} & \textbf{1.000E+00} \\
    \midrule
 
    \multirow{6}[0]{*}{\thead{TorchSig- \\ Real}} & VAE   & 8.085E-01 & \underline{8.631E-01} & 4.537E-01 & 9.792E-01 & 4.051E-01 & 1.506E-01 & 3.114E-01 & 1.679E-03 & 4.521E-01 \\
          & $\beta$-$\mathrm{VAE}_H$ & \underline{9.312E-01} & \textbf{8.712E-01} & 4.462E-01 & 9.695E-01 & 4.060E-01 & 1.173E-01 & 2.869E-01 & 8.663E-03 & 4.568E-01 \\
          & $\beta$-$\mathrm{VAE}_B$ & 7.477E-01 & 8.056E-01 & \underline{5.813E-01} & 9.161E-01 & 4.506E-01 & 9.251E-02 & 2.688E-01 & 1.498E-02 & 3.835E-01 \\
          & FactorVAE & 7.601E-01 & 8.013E-01 & 4.441E-01 & \textbf{9.870E-01} & 4.480E-01 & 1.609E-01 & 2.842E-01 & 1.692E-02 & 4.488E-01 \\
          & $\beta$-TCVAE & 8.336E-01 & 8.069E-01 & 4.269E-01 & 9.732E-01 & \underline{4.715E-01} & \underline{1.865E-01} & \underline{2.574E-01} & \underline{1.824E-02} & \underline{5.199E-01} \\
          & \textbf{Proposed} & \textbf{1.000E+00} & 3.503E-01 & \textbf{9.353E-01} & \underline{9.815E-01} & \textbf{4.813E-01} & \textbf{4.687E-01} & \textbf{2.648E-01} & \textbf{4.779E-02} & \textbf{9.868E-01} \\
    \bottomrule
    \end{tabular}%
  \label{tab:comparison}%
\end{table*}%

As we differentiate factors using different dimensions of representation, analogous to VAE-based models \cite{wang_Disentangled_2024}, we compared our approach with several VAE-based DRL models, including VAE~\cite{kingma_auto_2022}, $\beta$-VAE$_H$~\cite{higgins_beta_2017}, $\beta$-VAE$_B$~\cite{burgess_understanding_2018}, FactorVAE~\cite{kim_disentangling_2019}, and $\beta$-TCVAE~\cite{chen_isolating_2019}. Table~\ref{tab:comparison} summarizes the representation results of the proposed method and the aforementioned DRL models across three datasets. For each dataset, the maximum value of each metric is highlighted in bold, while the second-highest value is underlined.

Following the principle priorities discussed in Section~\ref{subsec:DRL}, we first focus on the metrics primarily reflecting \textit{explicitness}. On the POWDER dataset, the proposed method achieves the best Explicitness score of 0.9994, while the Explicitness scores of other methods remain around 0.1. On the WiSig-ManySig and TorchSig-Real datasets, the proposed method obtains Explicitness scores of 1.0 and 0.9868, respectively, whereas the other methods achieve scores around 0.6 and 0.5. These results indicate that the proposed method consistently performs exceptionally well in terms of \textit{explicitness}, a trend also observed in other metrics incorporating \textit{explicitness}.

In terms of the Modularity score, the comparison methods exhibit results that are closer to those of the proposed method. Across all three datasets, classical DRL methods achieve high Modularity scores, while the proposed method also obtains scores around 0.9, demonstrating its reliability in capturing \textit{modularity}. The performance on both the Explicitness and Modularity scores, suggests that the representations learned by traditional DRL methods may fail to capture all relevant factors comprehensively. In contrast, the proposed method not only maintains strong independence between factor representations but also ensures that each representation encodes complete and meaningful information.

We further evaluate the overall disentanglement using composite metrics including IRS, DCIMIG, JEMMIG, and SAP, which jointly reflect aspects of \textit{explicitness}, \textit{modularity}, and \textit{compactness}. On all three datasets, the proposed method achieves either the best or second-best performance across these metrics. Given the previously established advantages in \textit{explicitness} and \textit{modularity}, these results imply that the proposed method also exhibits strong \textit{compactness}, meaning that each factor influences only a limited and well-defined subset of the representation space. Although these metrics are theoretically designed to capture all three properties, their sensitivity to each aspect varies~\cite{carbonneau_Measuring_2022}. This further underscores the advantage of the proposed method, which consistently outperforms VAE-based DRL baselines from multiple measurement perspectives.

$Z_{\text{diff}}$, $Z_{\text{min}}$, and $Z_{\text{max}}$ are metrics proposed to evaluate how representations respond to controlled interventions. The proposed method achieves ideal performance on both $Z_{\text{diff}}$ and $Z_{\text{max}}$, with scores either reaching the upper bound of 1.0 or significantly outperforming other methods. This demonstrates that the learned representations can accurately reflect specific factors when corresponding components of the latent space are perturbed.

However, the proposed method performs poorly on $Z_{\text{min}}$. This outcome is expected, as the formulation of $Z_{\text{min}}$ does not theoretically guarantee higher scores for representations that exhibit desirable disentangling properties~\cite{Sepliarskaia_How_2021}. Similar behavior has also been observed with other metrics such as $Z_{\text{diff}}$ and SAP~\cite{Sepliarskaia_How_2021}. This underscores a broader challenge in the field of DRL: existing metrics do not always reliably capture the true quality or effectiveness of learned representations. To provide a more comprehensive evaluation, we further assess the learned representations through their performance on downstream tasks.

\begin{table}[t]
  \centering
  \renewcommand{\arraystretch}{1}
  \tabcolsep=.05cm
  \caption{Attribute Prediction Accuracy Comparison}
    \begin{tabular}{crrrrrrrrrr}
    \toprule
    \multirow{1}{*}{\thead{\\ Model}}
    &\multicolumn{3}{c}{POWDER}&\multicolumn{3}{c}{WiSig-ManySig}&\multicolumn{3}{c}{TorchSig-Real} & 
    \cr
    \cmidrule(lr){2-4} \cmidrule(lr){5-7} \cmidrule(lr){8-10}
    & \multicolumn{1}{c}{$f_{P}^{1}$} & \multicolumn{1}{c}{$f_{P}^{2}$} & \multicolumn{1}{c}{$f_{P}^{3}$} & \multicolumn{1}{c}{$f_{W}^{1}$} & \multicolumn{1}{c}{$f_{W}^{2}$} & \multicolumn{1}{c}{$f_{W}^{3}$} & \multicolumn{1}{c}{$f_{T}^{1}$} & \multicolumn{1}{c}{$f_{T}^{2}$} & \multicolumn{1}{c}{$f_{T}^{3}$} & \multicolumn{1}{c}{\thead{Avg}}
    \cr
    \midrule
    VAE   & 50.23  & 39.08  & 25.07  & 34.74  & 54.05  & 84.77  & 57.17  & 40.89  & 18.21  & 44.91  \\
    $\beta$-$\mathrm{VAE}_H$ & 50.23  & 39.08  & 25.07  & 30.85  & 45.30  & 79.28  & 58.22  & 43.84  & 16.60  & 43.16  \\
    $\beta$-$\mathrm{VAE}_B$ & 50.23  & 39.08  & 25.07  & 38.21  & 57.16  & 89.51  & 42.76  & 39.14  & 19.88  & 44.56  \\
    FactorVAE & 49.89  & 38.98  & 25.19  & 37.84  & 59.85  & 87.42  & 57.99  & 43.69  & 21.25  & 46.90  \\
    $\beta$-TCVAE & 50.23  & 39.08  & 25.07  & 37.35  & 58.55  & 88.62  & 59.98  & 43.77  & 24.00  & 47.41  \\
    \midrule
    Proposed & \textbf{99.44}  & \textbf{97.77}  & \textbf{99.98} & \textbf{99.22}  & \textbf{99.83}  & \textbf{99.73}  & \underline{99.62}  & \textbf{92.36}  & \textbf{86.07}  & \textbf{97.11}  \\
    \midrule
    No $\mathcal{L}_C$ & \underline{72.04}  & \underline{62.60}  & \underline{82.37}  & \underline{60.43}  & \underline{95.92}  & \underline{97.89}  & \textbf{99.94}  & \underline{73.68}  & \underline{38.95}  & \underline{75.98}  \\
    \bottomrule
    \end{tabular}%
  \label{tab:comparison2}%
\end{table}%

Another widely adopted metric for evaluating how effectively a representation captures an underlying factor is the Attribute Prediction Accuracy (APA)~\cite{Lin_Improving_2019}. This metric assesses the predictive utility of learned representations by splitting the dataset into training and testing subsets. Representations from the training set are used to train a LinearSVM classifier in a supervised manner, and the classification accuracy on the testing set is reported for each factor.

Table~\ref{tab:comparison2} presents the APA results for the proposed DRL method across different datasets, using an 8:2 train-test split. Notably, even without introducing the supervised loss $\mathcal{L}_{CE}$, the APA scores obtained by the proposed method consistently outperform those of existing DRL models. This demonstrates that the learned representations successfully isolate individual factors. In particular, this supports the theoretical assertion that $\nabla_{x}\log p(z^c|x)$ serves as a reliable approximation to $\nabla_{x}\log p(f^c|x)$, as discussed in Section~\ref{subsec:DRL}.

\begin{figure*}[t]
  \centering
  \subfloat[DRL's representation $\mathcal{Z}^c$ on POWDER]{\includegraphics[width=.63\columnwidth]{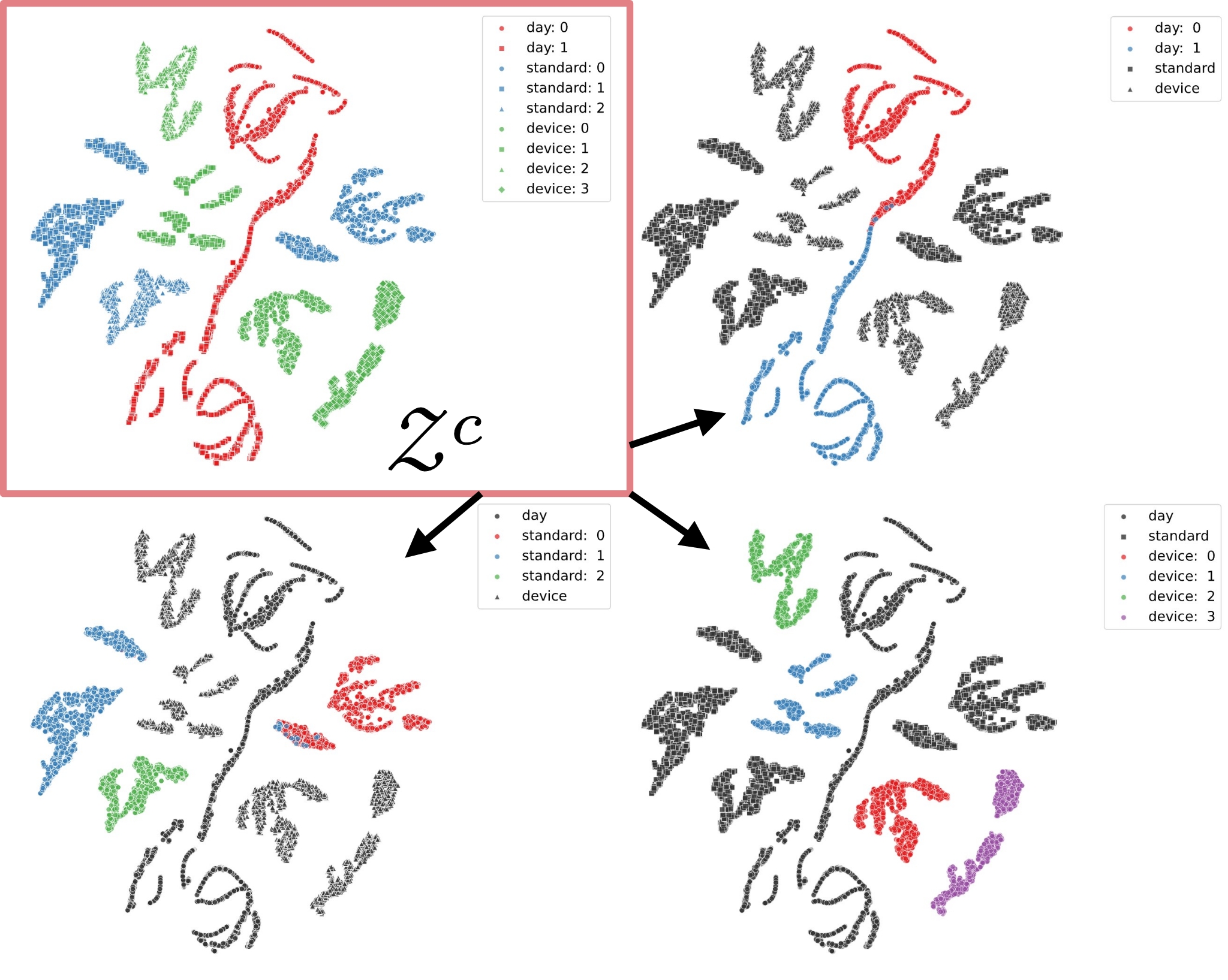}\hspace{5mm}\label{fig:tsne_powder}}
  \subfloat[DRL's representation $\mathcal{Z}^c$ on WiSig-Manysig]{\includegraphics[width=.63\columnwidth]{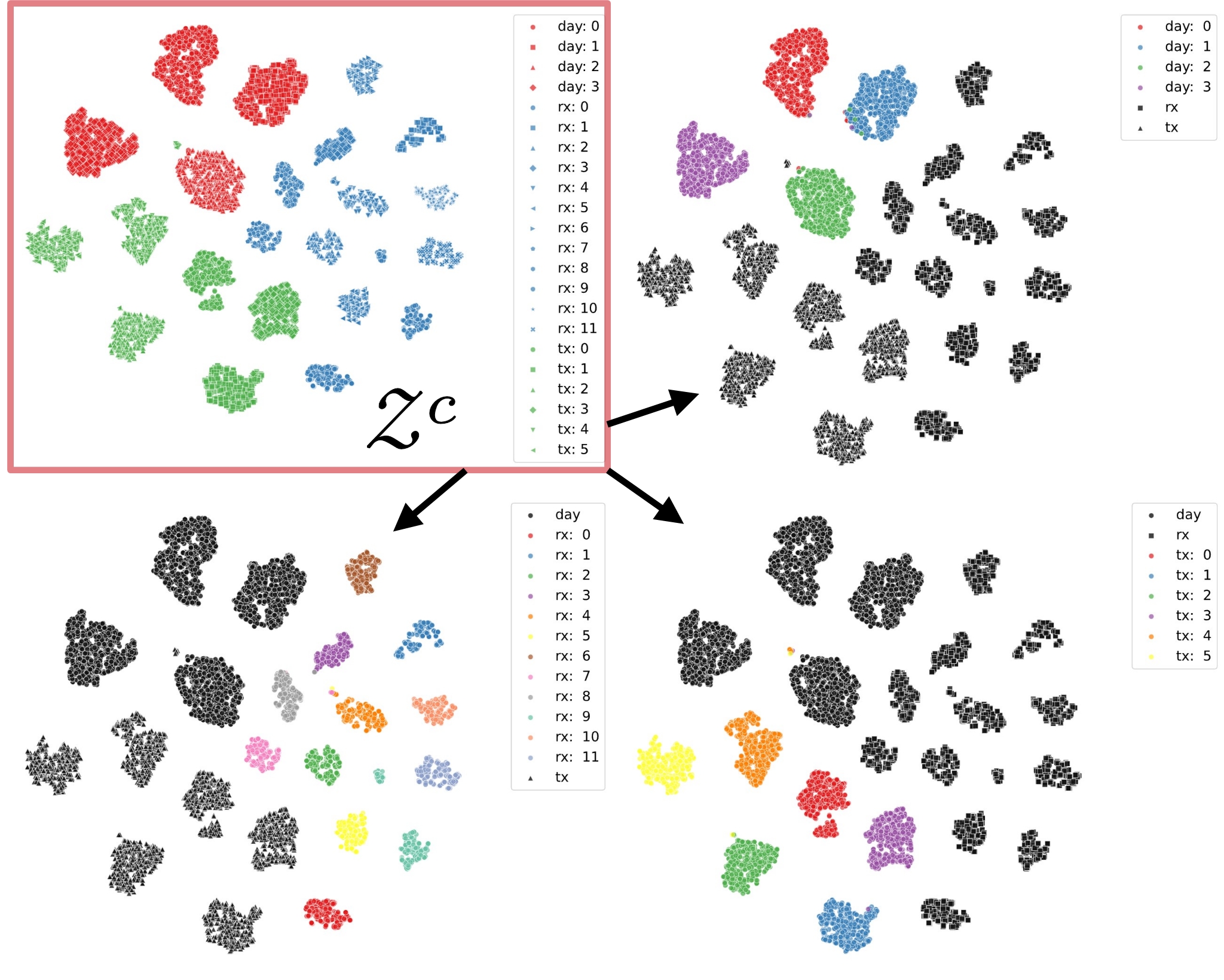}\hspace{5mm}\label{fig:tsne_wisig}}
  \subfloat[DRL's representation $\mathcal{Z}^c$ on TorchSig-Real]{\includegraphics[width=.63\columnwidth]{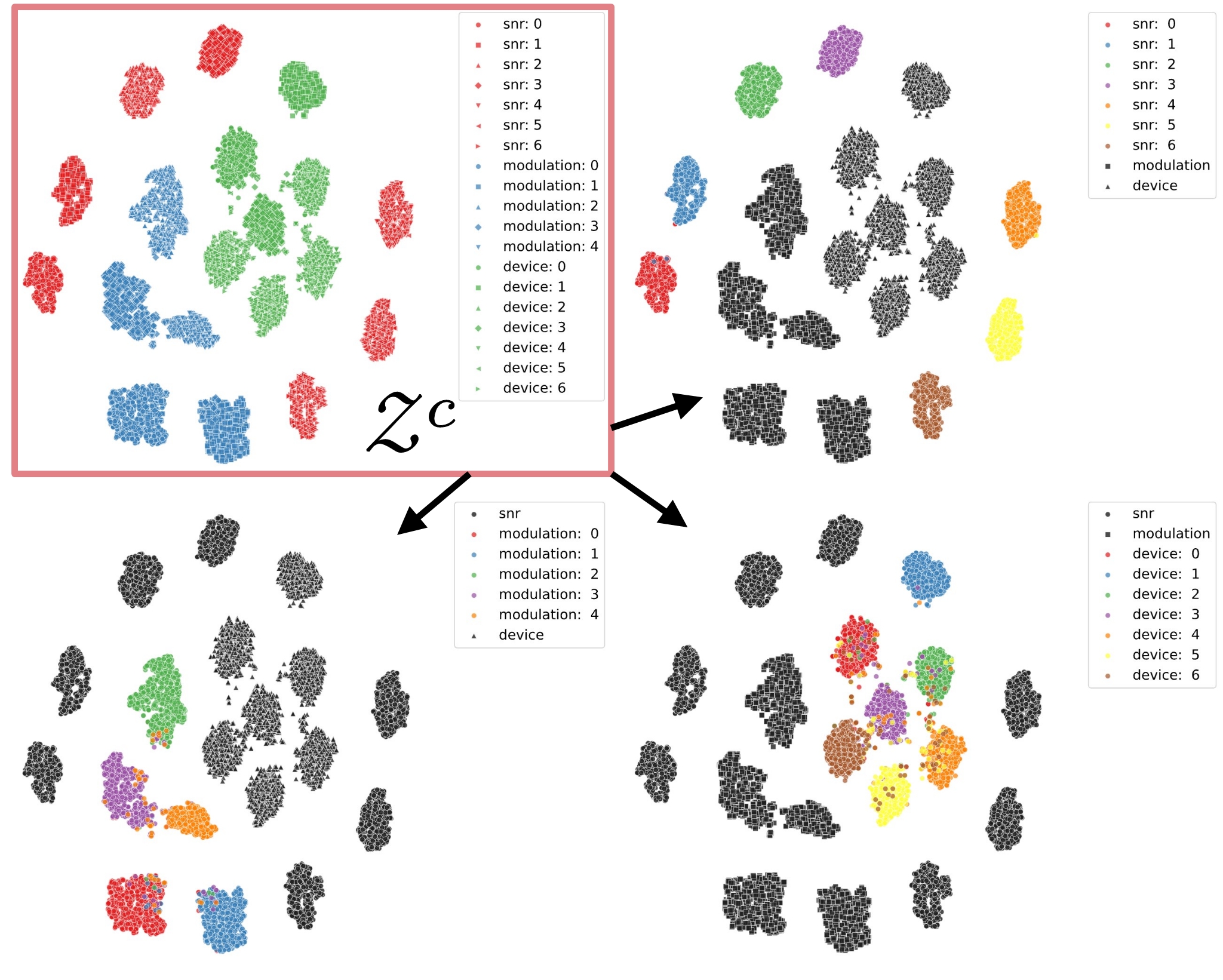}\hspace{5mm}\label{fig:tsne_TorchSig-Real}}\\

  \subfloat[Common classifier's feature space POWDER]{\includegraphics[width=.63\columnwidth]{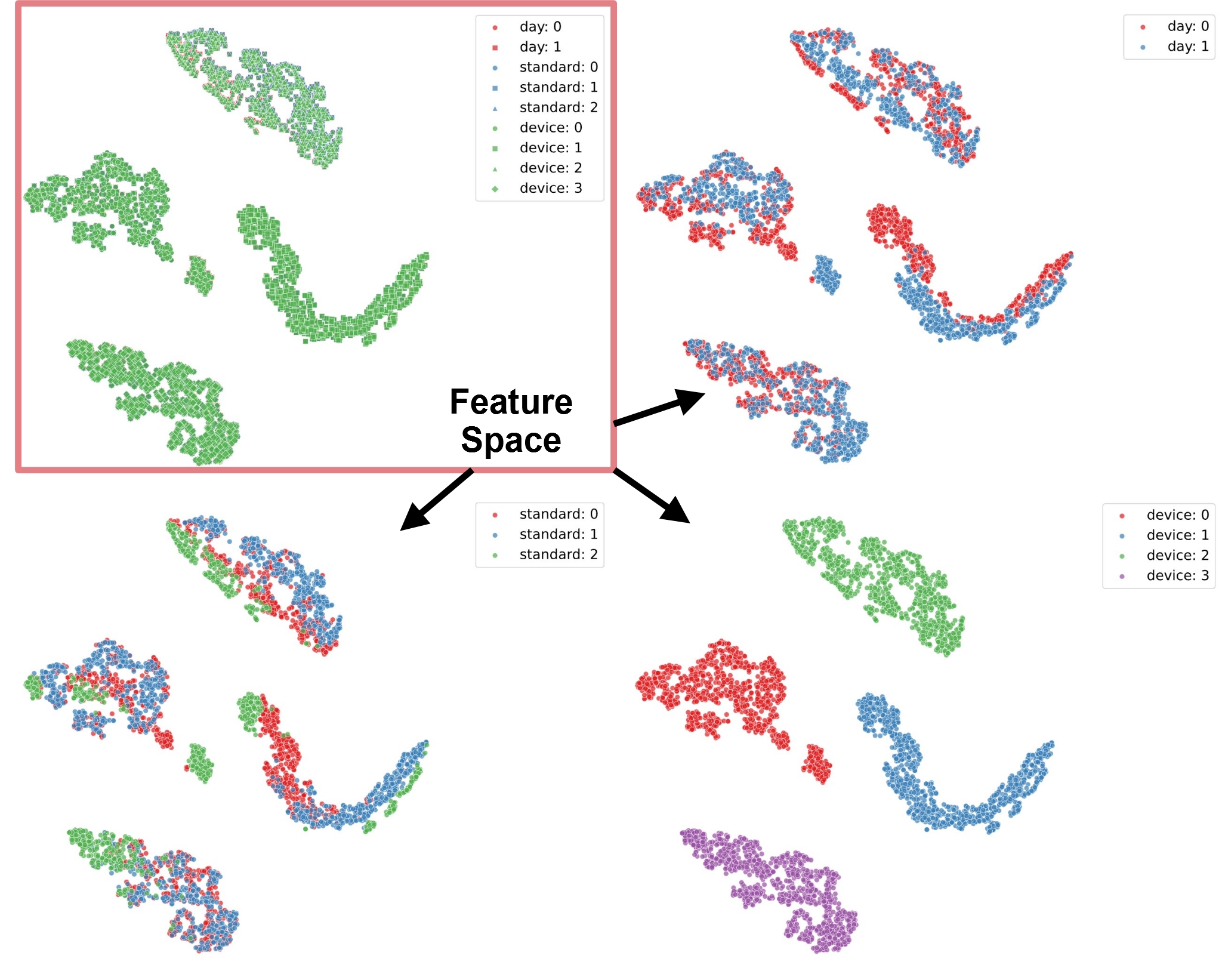}\hspace{5mm}\label{fig:tsne_powder_seperate}}
  \subfloat[Common classifier's feature space WiSig-Manysig]{\includegraphics[width=.63\columnwidth]{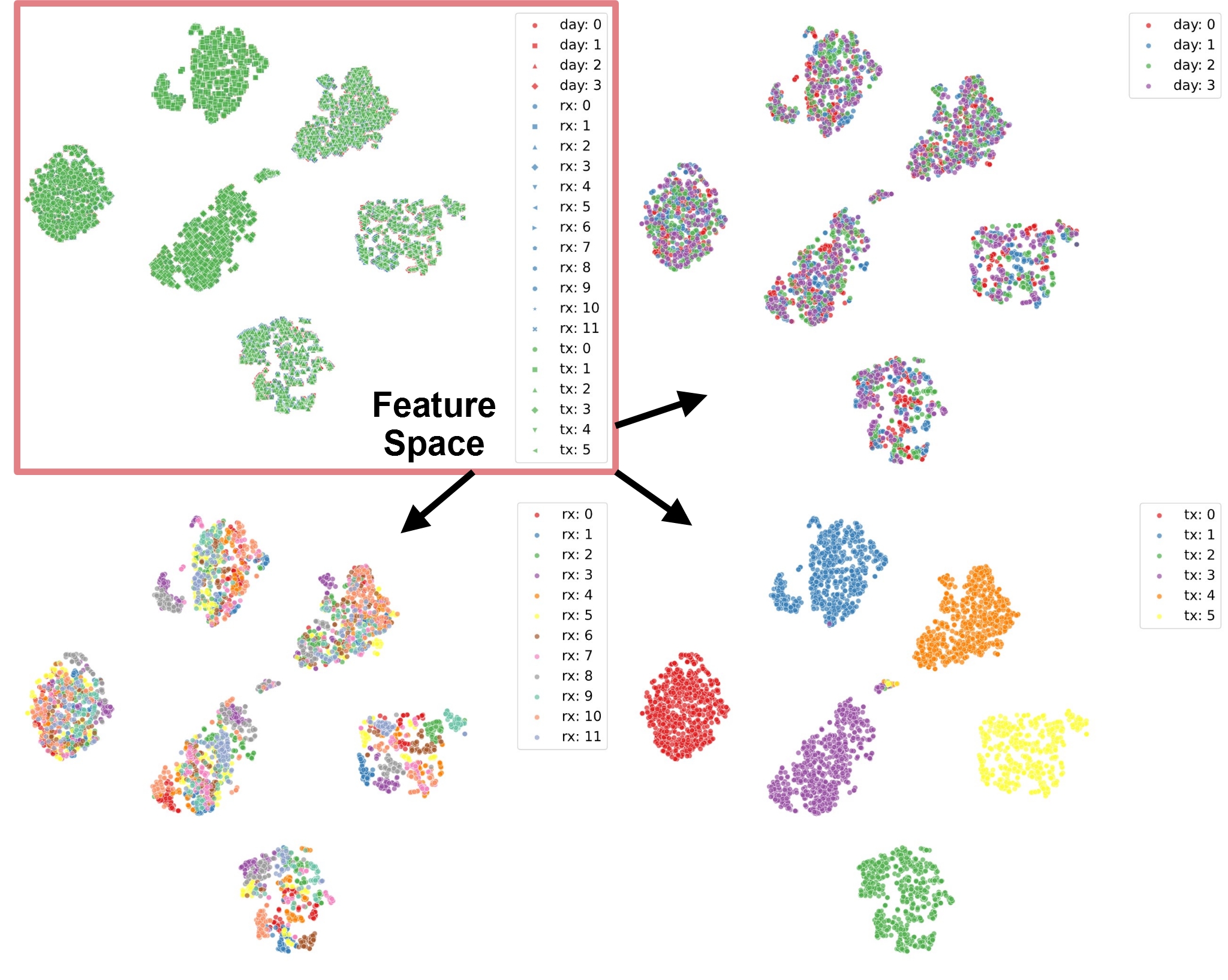}\hspace{5mm}\label{fig:tsne_wisig_seperate}}
  \subfloat[Common classifier's feature space TorchSig-Real]{\includegraphics[width=.63\columnwidth]{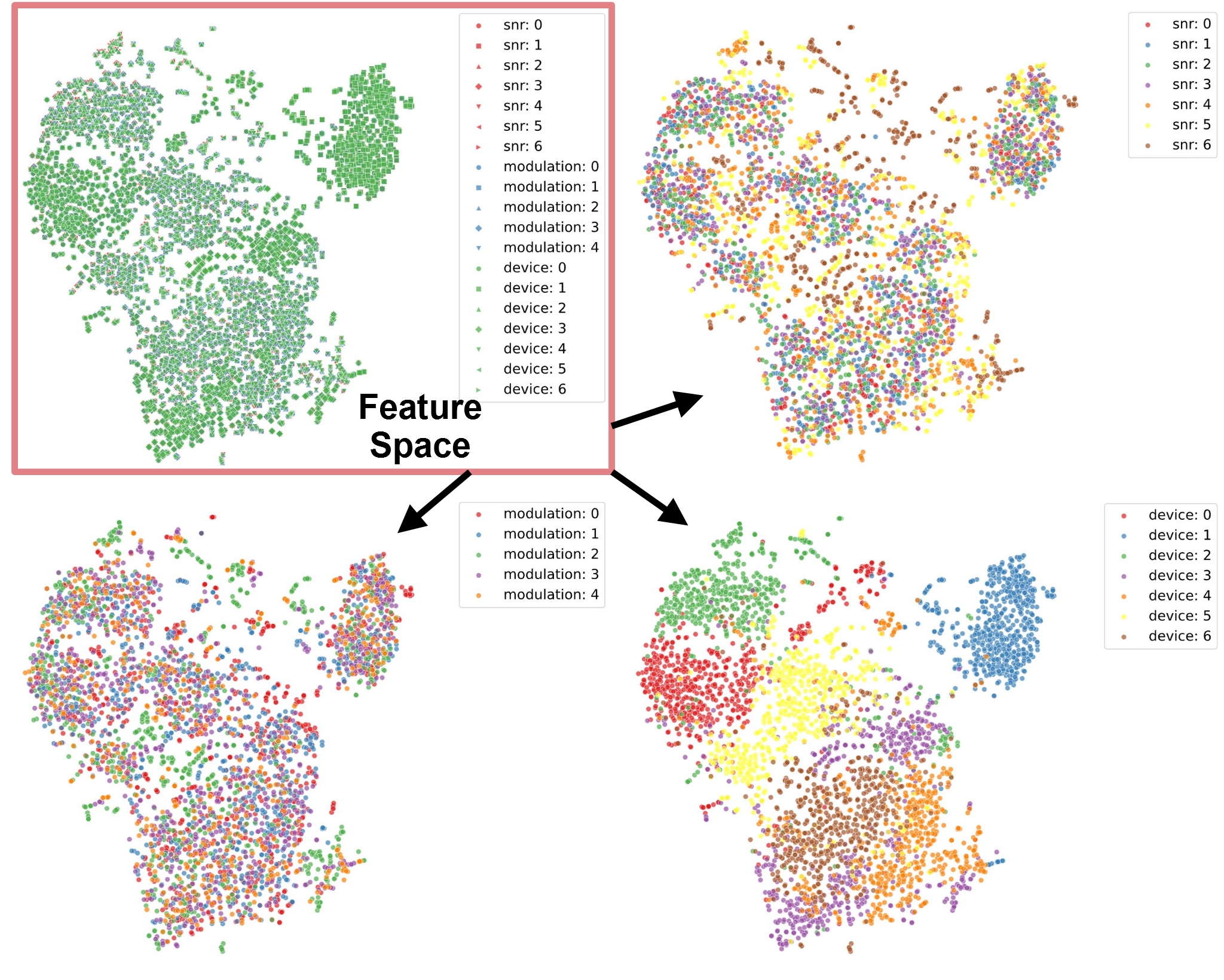}\hspace{5mm}\label{fig:tsne_TorchSig-Real_seperate}}
  \caption{Diagram and phenomenon explanation of DRL. 
  (a) shows the Probabilistic Graphical Model (PGM) composed of factor, signal, and representation. (b) illustrates the widely acknowledged principles of modularity and explicitness in DRL. (c) demonstrates the feature space distribution of signal samples \ding{172} $\sim$ \ding{178} in (d) using RFF, Modulation, and Receiver ID as examples. The overlapping surfaces formed by different factors result in signal samples with distinct representations.}\label{fig:tsne_proposed}
\end{figure*}

\subsubsection{Visualization of the Representation}\label{subsubsec:visualization}

To better understand how the proposed DRL framework achieves disentanglement and to highlight its differences from conventional classification models, we extract the factor representations $\mathcal{Z}^{c}$ with ResNet-18 as the feature extractor $E$ across all three datasets. For comparison, we also extract the latent feature spaces from a standard classification model composed of a ResNet-18 backbone followed by a classification head, trained to perform transmitter (RFF) classification.

Fig. \ref{fig:tsne_powder}, \ref{fig:tsne_wisig}, \ref{fig:tsne_TorchSig-Real} and Fig. \ref{fig:tsne_powder_seperate}, \ref{fig:tsne_wisig_seperate}, \ref{fig:tsne_TorchSig-Real_seperate} visualize $\mathcal{Z}^{c}$ and the latent features using t-distributed Stochastic Neighbor Embedding (t-SNE) plots \cite{vab_Visualizing_2008}, respectively, showing the flow of factors after feature extraction. In Fig. \ref{fig:tsne_powder}, \ref{fig:tsne_wisig}, \ref{fig:tsne_TorchSig-Real}, the upper-left corner shows that $\mathcal{Z}^{c}$ distinguishes three different factors, whereas in Fig. \ref{fig:tsne_powder_seperate}, \ref{fig:tsne_wisig_seperate}, \ref{fig:tsne_TorchSig-Real_seperate}, the upper-left corner focuses solely on transmitters (RFF). This means that, for an input signal, when entering the DRL framework, three coordinates are obtained in the representation space $\mathcal{Z}^{c}$, with each coordinate corresponding to a factor.

\begin{table}[t]
  \centering
  \renewcommand{\arraystretch}{1}
  \tabcolsep=.1cm
  \caption{Ablation Study of the Proposed Losses for Disentanglement}
    \begin{tabular}{cccccccc}
    \toprule
    \multirow{2}[0]{*}{} & \multicolumn{1}{c}{\multirow{2}[0]{*}{$\mathcal{L}_{C}$}} & \multicolumn{1}{c}{\multirow{2}[0]{*}{$\mathcal{L}_{FD}$}} & \multirow{2}[0]{*}{$\mathcal{L}_{SE}$} & \multirow{2}[0]{*}{$\mathcal{L}_{RC}$} & \multicolumn{3}{c}{DCIMIG}
    \cr
    \cmidrule{6-8}
          &       &       &       &       & POWDER & WiSig & TorchSig-Real
    \cr
    \midrule
    Proposed & \multicolumn{1}{c}{\ding{51}} & \multicolumn{1}{c}{\ding{51}} & \ding{51} & \ding{51} & \textbf{0.7528} & \textbf{0.4854} & \textbf{0.4687} \\
    \midrule
    \multirow{3}[0]{*}{With $\mathcal{L}_{C}$} & \multicolumn{1}{c}{\ding{51}} & \multicolumn{1}{c}{\ding{51}} & \ding{51} & \textcolor{lightgray}{\ding{55}}      & 0.7495  & 0.4512  & 0.4520  \\
          & \multicolumn{1}{c}{\ding{51}} & \multicolumn{1}{c}{\ding{51}} & \textcolor{lightgray}{\ding{55}}      & \ding{51} & 0.7106  & 0.4609  & 0.4580  \\
          & \multicolumn{1}{c}{\ding{51}} & \textcolor{lightgray}{\ding{55}}      & \ding{51} & \ding{51} & 0.7256  & 0.4670  & 0.4542  \\
    \midrule
    \multirow{4}[0]{*}{No $\mathcal{L}_{C}$} & \textcolor{lightgray}{\ding{55}}      & \multicolumn{1}{c}{\ding{51}} & \ding{51} & \ding{51} & \textbf{0.2864} & \textbf{0.2151} & \textbf{0.2158} \\
          & \textcolor{lightgray}{\ding{55}}      & \multicolumn{1}{c}{\ding{51}} & \ding{51} & \textcolor{lightgray}{\ding{55}}      & 0.1827  & 0.0687  & 0.0556  \\
          & \textcolor{lightgray}{\ding{55}}      & \multicolumn{1}{c}{\ding{51}} & \textcolor{lightgray}{\ding{55}}      & \ding{51} & 0.2343  & 0.1889  & 0.1990  \\
          & \textcolor{lightgray}{\ding{55}}      & \textcolor{lightgray}{\ding{55}}      & \ding{51} & \ding{51} & 0.2148  & 0.1435  & 0.1668  \\
    \bottomrule
    \end{tabular}%
  \label{tab:ablation}%
\end{table}%

This results in significant differences between the proposed DRL framework and ResNet in the t-SNE plots reflected by the labels in the lower-left, upper-right, and lower-right corners. For example, in Fig. \ref{fig:tsne_TorchSig-Real} and \ref{fig:tsne_TorchSig-Real_seperate}, using the results from the TorchSig-Real dataset, the proposed DRL framework clearly separates the representation space into SNR, modulation, and transmitter (RFF), and classification is performed within each factor. The classification task for each factor only uses a subpart of the representation $\mathcal{Z}^{c}$, thus enabling classification in a more compact space. In contrast, the loose representation space of ResNet makes it unable to distinguish SNR and modulation, which weakens its ability to distinguish transmitters.

Looking back at Fig. \ref{fig:exp_mod_com}, the proposed DRL framework satisfies the requirements \textit{explicitness} (finding all factors), \textit{modularity} (independence between factors), and \textit{compactness} (a sufficiently compact representation space $\mathcal{Z}^{c}$).



\subsubsection{Ablation Study}



An ablation study is conducted to evaluate the contribution of each component of the proposed method and the corresponding loss functions to the model's performance.

As shown in Table \ref{tab:ablation}, removing different loss functions led to varying degrees of performance degradation, measured by the DCIMIG metric. Among them, the classification loss $\mathcal{L}_{C}$ had the most significant impact, reducing the DCIMIG score from approximately 0.7 to 0.2. This highlights its critical role in introducing inductive bias. The other three loss functions, $\mathcal{L}_{FD}$, $\mathcal{L}_{SE}$, and $\mathcal{L}_{RC}$, each contributed positively to performance, regardless of whether $\mathcal{L}_{C}$ was applied. Specifically, $\mathcal{L}_{RC}$ had the greatest additional effect, followed by $\mathcal{L}_{FD}$ and then $\mathcal{L}_{SE}$.

These results are consistent with the interpretations in Section \ref{subsec:structure}, where $\mathcal{L}_{RC}$ primarily enhances \textit{explicitness}, $\mathcal{L}_{FD}$ supports \textit{modularity}, and $\mathcal{L}_{SE}$ contributes to \textit{compactness}. The differing degrees of performance degradation reflect the relative importance of these properties.

\subsection{Classification Performance}\label{subsec:classification}

 Table \ref{tab:classification} presents the classification results on different factors across three datasets under an 8:2 training-testing split. We compare the proposed model using only the classification loss $\mathcal{L}_C$ (implemented as $1 \times$ (1 feature extractor + 3 classifiers)) with seperate classifiers that directly uses ResNet-18 ($3 \times$ (1 feature extractor + 1 classifier)). On all three datasets, the proposed method achieves the highest classification accuracy. Moreover, on the TorchSig-Real dataset, which features more challenging SNR conditions, the proposed method significantly outperforms the other two approaches in both modulation and RFF classification. This indicates that the proposed model successfully disentangles the effects of SNR in the representation space and better utilizes the representation capacity for modulation and RFF factors.

\begin{figure*}[!t]
  \centering
  {\includegraphics[width=2\columnwidth]{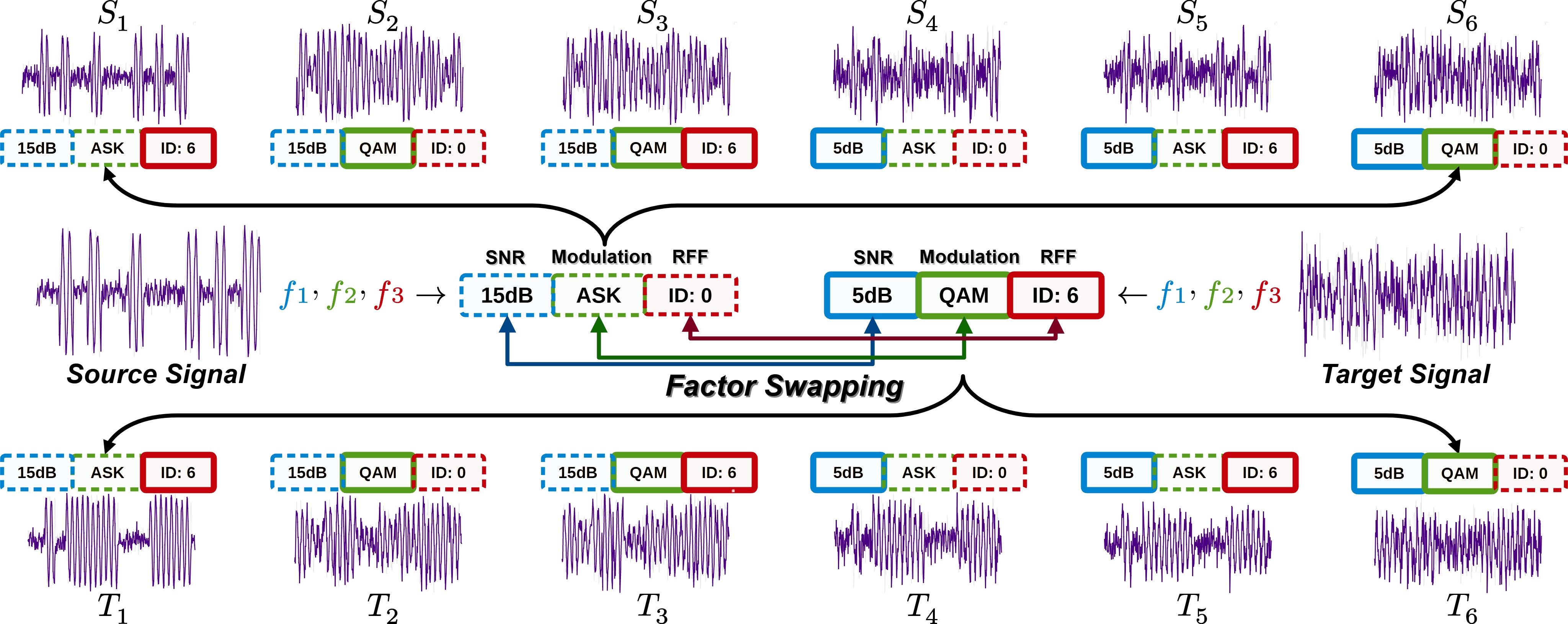}\hspace{5mm}
  \caption{The conditioned generated signal obtained by swapping the factors of the source signal and the target signal.}\label{fig:swap}}
\end{figure*}

\begin{table}[t]
  \centering
  \renewcommand{\arraystretch}{1}
  \tabcolsep=.15cm
  \caption{Classification Performance Comparison}
    \begin{tabular}{cccccc}
    \toprule
    Dataset & Method & $f^1$ & $f^2$ & $f^3$ & Avg \\
    \midrule
    \multirow{3}[0]{*}{POWDER} & \textbf{Proposed} & \textbf{98.51} & \textbf{94.98} & \textbf{99.99} & \textbf{97.83} \\
          & $\mathcal{L}_{C}$ only & 96.61  & 93.47  & \underline{99.97}  & 96.68  \\
          & Separate Classifier & \underline{97.67}  & \underline{93.99}  & 99.61  & \underline{97.09}  \\
    \midrule
    \multirow{3}[0]{*}{\thead{WiSig-ManySig}} & \textbf{Proposed} & \textbf{98.54} & \textbf{99.71} & \textbf{99.68} & \textbf{99.31} \\
          & $\mathcal{L}_{C}$ only & \underline{96.43}  & 99.32  & 99.42  & 98.39  \\
          & Separate Classifier & 95.65  & \underline{99.38}  & \underline{99.46}  & \underline{98.16}  \\
    \midrule
    \multirow{3}[0]{*}{TorchSig-Real} & \textbf{Proposed} & \textbf{99.18} & \textbf{81.46} & \textbf{66.27} & \textbf{82.30} \\
          & $\mathcal{L}_{C}$ only & 96.88  & 75.55  & 42.98  & 71.80  \\
          & Separate Classifier & \underline{98.60}  & \underline{77.82}  & \underline{45.23}  & \underline{73.88}  \\
    \bottomrule
    \end{tabular}%
  \label{tab:classification}%
\end{table}%

\subsection{Generalization Performace}\label{subsec:generalization}

\subsubsection{Representation Swapping}\label{subsubsec:swapping}


\begin{figure}[t]
  \centering
  \subfloat[ID: 0 $\rightarrow$ ID: 6]{\includegraphics[height=.45\columnwidth]{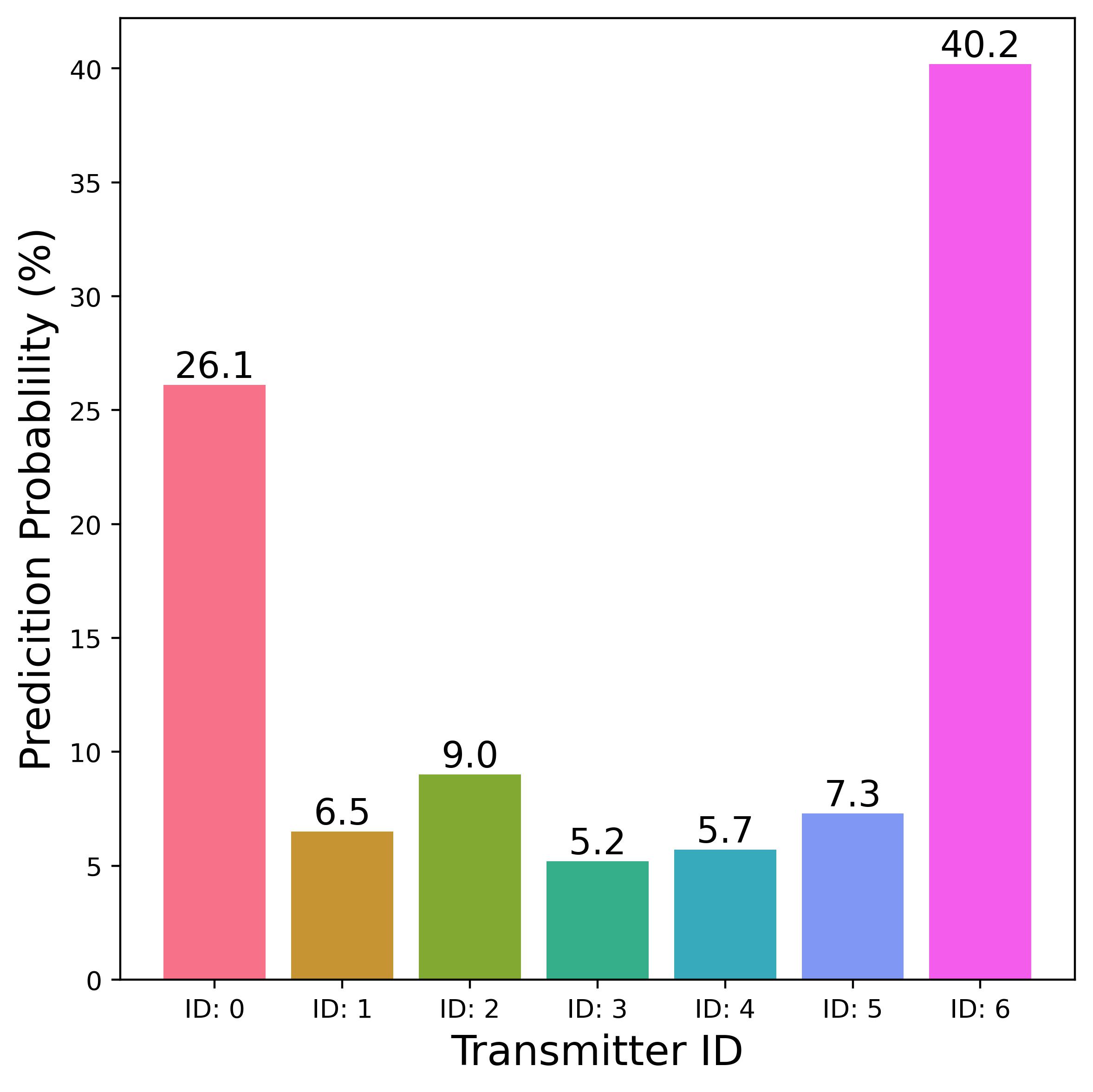}\hspace{2mm}\label{fig:bar_plot_S}}
  \subfloat[ID: 6 $\rightarrow$ ID: 0]{\includegraphics[height=.45\columnwidth]{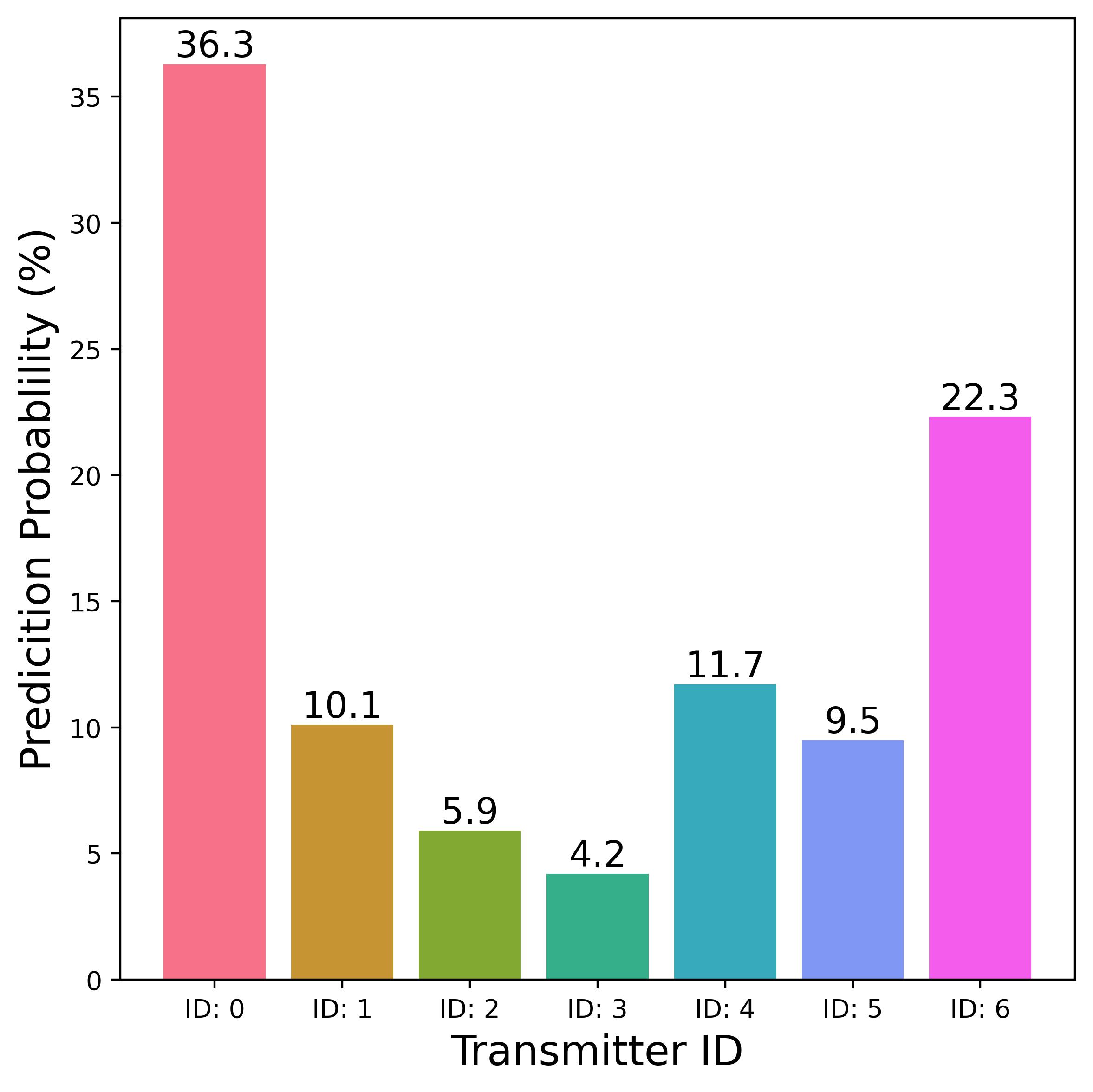}\label{fig:bar_plot_T}}
  \caption{Average predicted probability of generated signals after their representation of Transmitter/RFF swapped from (ID:0, ID:6) to (ID:6, ID:0).}\label{fig:swap_rff}
\end{figure}

The disentangled representation $\mathcal{Z}^c$ learned by proposed model not only supports classification tasks, but also enables factor-based conditional signal generation. As shown in Fig. \ref{fig:swap}, the model performs \textit{factor swapping} between two input signals, recombining their representations to generate new signals with different combinations of underlying factors. The source and target signals differ in all annotated factors, including SNR (5\,dB, 15\,dB), modulation scheme (ASK (\textless 4 symbol), QAM (\textgreater 16 symbol)), and RFF ID (ID:0, ID:6). 

The model first disentangles both signals into their representations $\mathcal{Z}$ and then obtains a condition factor map $M_f$. By permutating the factors in $z^c$ while excluding the original two representations, we obtain six unique factor combinations. Since the input to the generator $G$ is the noised version $x_t$ of both the source and target signals under a diffusion process, a total of 12 generated signals are obtained. In Fig. \ref{fig:swap}, the upper half shows the signals $S_1 \sim S_6$ generated from the source signal, and the lower half shows $T_1 \sim T_6$ generated from the target signal. 

All 12 generated signals exhibit distinct characteristics aligned with their underlying factors. For instance, $S_1 \sim S_3$ and $T_1 \sim T_3$ are conditioned on a high-SNR $z^c$ (15\,dB), while $S_4 \sim S_6$ and $T_4 \sim T_6$ correspond to a low-SNR $z^c$ (5\,dB). A visual comparison clearly shows that $S_4 \sim S_6$ and $T_4 \sim T_6$ have higher noise levels than $S_1 \sim S_3$ and $T_1 \sim T_3$.

Regarding modulation schemes, signals $(S_1, S_4, S_5)$ and $(T_1, T_4, T_5)$ exhibit characteristics of ASK modulation, while $(S_2, S_3, S_6)$ and $(T_2, T_3, T_6)$ correspond to QAM. The amplitude distributions of $(S_1, S_4, S_5)$ and $(T_1, T_4, T_5)$ are more concentrated, while $(S_2, S_3, S_6)$ and $(T_2, T_3, T_6)$ cover a broader amplitude range and show sudden phase changes—hallmark features of QAM compared to ASK.

As the Transmitter (RFF) factor cannot be easily observabled through visual inspection, a quantitative evaluation of signal representations in Fig. \ref{fig:swap_rff} is processed to assess how well the proposed method captures RFF-specific features. The bar plots in Fig. \ref{fig:bar_plot_S} and \ref{fig:bar_plot_T} show all samples of (ID:0, ID:6) in the dataset, along with the average predicted probabilities obtained after swapping them to (ID:6, ID:0), using a ResNet-18 classifier trained on the entire dataset. It can be observed that the swapped samples are classified as the Transmitter/RFF labels corresponding to their swapped Transmitter/RFF $z^c$.

\begin{figure*}[!t]
  \centering
  {\includegraphics[width=2\columnwidth]{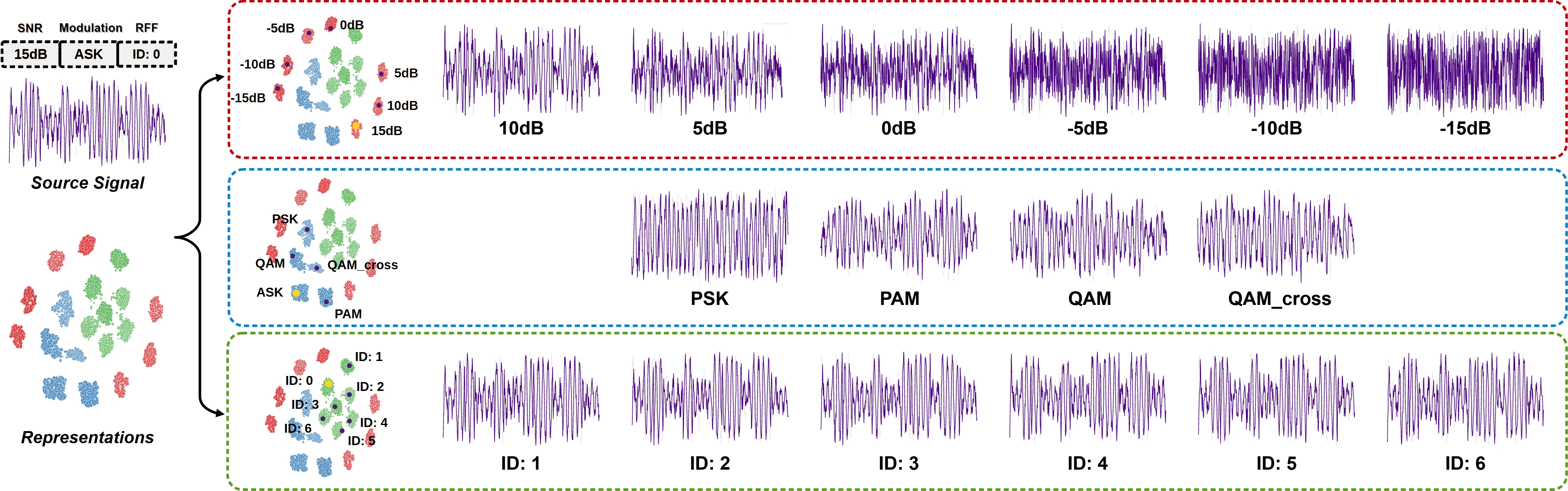}\hspace{5mm}
  \caption{The conditioned generated signal obtained by sampling factors in the representation space. }\label{fig:swap_all}}
\end{figure*}

In addition to expressing distinct factors, the signal generator $G$ also preserves partial characteristics of the original signal. For example, $S_1 \sim S_3$ resemble the waveform of the source signal but differ significantly from $T_1 \sim T_3$. This is because the diffusion model uses different inputs as $x_0$, leading to different unconditional features $p(x)$ in Eq. \eqref{eq:bayes}. Similarly, $T_2 \sim T_4$ resemble the waveform of the target signal, whereas $S_4 \sim S_6$ do not.

From another perspective, the symbol distribution of signals could also be considered a disentangleable factor. However, the proposed model does not explicitly disentangle digital symbol content. It is unable to interpret whether signals like $T_1$ or $S_3$ contain meaningful symbol sequences or how those symbols are generated, causing like $S_1$ differs from $T_1$.

\subsubsection{Representation Resampling}\label{subsubsec:traversal}

Fig. \ref{fig:swap_all} illustrates the conditioned generated signal obtained by sampling factors in the representation space. Since the model partitions the representation space based on different factors, representations can be sampled according to semantic information. In Fig. \ref{fig:swap_all}, the source signal samples a latent code $ z^c $ from the spaces corresponding to SNR, modulation, and RFF factors, and replaces its original $ z^c $ with the sampled one. 

After sampling the SNR factor, it can be observed that the generated signal exhibits different SNRs. After sampling the modulation factor, for instance, PSK, the generated signal shows uniform amplitude without the amplitude modulation characteristic of the source signal's ASK; meanwhile, for PAM, QAM, and cross QAM, the generated signals exhibit richer amplitude modulation and phase transitions. Sampling the RFF factor does not substantially alter the waveform of the source signal, which is consistent with the physical properties of real-world transmission: the RFF between different transmitters does not affect the modulation scheme of the transmitted signals and has minimal impact on the noise power.

\section{Discussion on Limitations and Future Expectations}\label{sec:discussion}

Despite the excellent performance in disentangling and supporting downstream tasks, the proposed framework still has the following limitations, which may arise in practical scenarios:

\begin{enumerate}

\item \textbf{Hyperparameter Settings}

Due to the introduction of disentangling-related losses, the number of independent losses is significantly higher compared to previous RFFI works. This means that the model has to balance more $\lambda$ weight parameters, which makes the settings more difficult. For example, for the TorchSig-Real dataset, tasks like modulation and RFF classification are much more challenging than SNR, meaning that the loss weights assigned to modulation and RFF should be larger. These weight settings typically require prior knowledge of the task and may only be reasonably set after some experimental trial and failure. 

In the future, for loss-guided DL models, automated weight allocation, or even training paradigms that do not require weight allocation, are urgently needed.

\item \textbf{Incorporation of Prior Knowledge}

In the disentangling process, whether label losses $\mathcal{L}_C$ are introduced as prior knowledge significantly influences the model's disentangling performance. This means that providing only the signal data without factors and labels is difficult for the model to determine which factors to extract (If there are three dimensions for representation, which dimension is for SNR/modulation/RFF?). The proposed method directly uses labels like to guide the model in achieving disentangling. However, in practical applications, the factors to be extracted are often far more than just three, and labels can be difficult to obtain. 

In the future, prior knowledge may be incorporated directly into the model with \textbf{causal representations}, or through the input of knowledge related to the \textbf{physical process} of signal generation, which may be more practical than relying on labels.
\end{enumerate}

\section{Conclusion}\label{sec:conclusion}

This paper presents a novel paradigm for RFFI, guided by the principles of Explicitness, Modularity, and Compactness incorporating a DRL framework. By integrating multiple loss functions for disentanglement, classification and generation tasks, the framework effectively disentangles different factors in the process of practicle signal generation into specific reprensentations, leading to interpretable feature extraction. The experiments on two public benchmark datasets and one collected dataset substantially demonstrate that the framework simultaneously enhances performance of RFFI and enables factor-based conditional signal generation, confirming the superiority of the proposed approach over conventional methods. These results underscore the application potential of DRL in RFF-based research. The proposed paradigm opens up promising directions for future RFFI research, such as applying DRL to branch the RFFI to more complex and dynamic wireless scenarios. Physical process-based disentanglement methods may also be a way to supplement interpretability and robustness.

\bibliographystyle{IEEEtran}
\bibliography{ref}

\end{document}